\documentclass[sigconf, 9pt]{acmart}

\usepackage[english]{babel}
\usepackage{blindtext}
\usepackage{tikz,pgfplots}
\usetikzlibrary{calc}
\usepackage{tkz-euclide}
\pgfplotsset{compat=1.5}
\usepgfplotslibrary{polar}
\usepackage{epstopdf} %converting EPS to PDF
\usepackage{amsmath}
\usepackage{lipsum}
\usepackage{wrapfig}
\usepackage{verbatim}
\usepackage{mathrsfs}
\usepackage{accents}
\usepackage{acronym}
\usepackage{graphicx}
\usepackage{textcomp}
\usepackage{xcolor}
\usepackage{mathtools}
\usepackage{color,soul}
\usepackage{graphicx}
\usepackage{booktabs}
\usepackage[font={footnotesize}]{caption}
\usepackage[font={footnotesize}]{subcaption}
\usepackage{algorithm}
\usepackage{algorithmic}
\usepackage{fancyhdr}

\DeclareMathOperator*{\argmin}{arg\,min}

\usepackage{caption}
\usepackage{subcaption}
\usepackage{cleveref}
\usepackage{url}

\newcommand{\name}{HiSAC}

\newcommand{\eq}[1]{Eq.~\eqref{#1}}
\newcommand{\fig}[1]{Fig.~\ref{#1}}
\newcommand{\tab}[1]{Tab.~\ref{#1}}
\newcommand{\secref}[1]{Section~\ref{#1}}
\newcommand{\alg}[1]{Alg.~\ref{#1}}

\acrodef{6g}[6G]{Sixth Generation}
\acrodef{5g}[5G-NR]{Fifth Generation-New Radio}
\acrodef{4g}[4G]{Fourth Generation}
\acrodef{3gpp}[3GPP]{$3^{\rm rd}$ Generation Partnership Project}
\acrodef{prop}[\textit{MIMORPH}]{MIMO Radio Platform for Heterogeneous wireless systems}

\acrodef{3gpp}[3GPP]{3rd Generation Partnership Project}
\acrodef{abft}[A-BFT]{Association Beamforming Training}
\acrodef{ack}[ACK]{Acknowledge}
\acrodef{adc}[ADC]{Analog-to-Digital Converter}
\acrodef{dac}[DAC]{Digital-to-Analog Converter}
\acrodef{aoa}[AoA]{Angle of Arrival}
\acrodef{aod}[AoD]{Angle of Departure}
\acrodef{ar}[AR]{Auto-Regressive}
\acrodef{ap}[AP]{Access Point}
\acrodef{amc}[AMC]{Advanced Mezzanine Card}
\acrodef{awv}[AWV]{Antenna Wave Vector}
\acrodef{axi}[AXI]{Advanced eXtensible Interface}
\acrodef{ar}[AR]{Auto-Regressive}
\acrodef{ber}[BER]{Bit Error Rate}
\acrodef{bft}[BFT]{Beamforming Training}
\acrodef{bp}[BP]{Beam Pattern}
\acrodef{bpsk}[BPSK]{Binary Phase-Shift Keying}
\acrodef{brp}[BRP]{Beam Refinement Phase}
\acrodef{bwp}[BWP]{Bandwidth Part}
\acrodef{bs}[BS]{Base Station}
\acrodef{cs}[CS]{Compressed Sensing}
\acrodef{cdf}[CDF]{Cumulative Distribution Function}
\acrodef{cef}[CEF]{Channel Estimation Field}
\acrodef{cacfar}[CA-CFAR]{Cell-Averaging Constant False-Alarm Rate}
\acrodef{cacc}[CACC]{Cross-Antenna Cross-Correlation}
\acrodef{ca}[CA]{Carrier Aggregation}
\acrodef{casr}[CASR]{Cross-Antenna Signal Ratio}
\acrodef{csi-rs}[CSI-RS]{CSI-Reference Signal}
\acrodef{cfo}[CFO]{Carrier Frequency Offset}
\acrodef{cir}[CIR]{Channel Impulse Response}
\acrodef{cfr}[CFR]{Channel Frequency Response}
\acrodef{csi}[CSI]{Channel State Information}
\acrodef{csirs}[CSI-RS]{CSI-Reference Signal}
\acrodef{cv}[CV]{Constant Velocity}
\acrodef{cnn}[CNN]{Convolutional Neural Network}
\acrodef{cots}[COTS]{Commercial-Off-The-Shelf}
\acrodef{dft}[DFT]{Discrete Fourier Transform}
\acrodef{dl}[DL]{Deep Learning}
\acrodef{dma}[DMA]{Direct Memory Access}
\acrodef{dmg}[DMG]{Directional Multi Gigabit}
\acrodef{dm-rs}[DM-RS]{De-Modulation-Reference Signal}
\acrodef{dti}[DTI]{Data Transfer Interval}
\acrodef{dft}[DFT]{Discrete Fourier Transform}
\acrodef{dtft}[DTFT]{Discrete-Time Fourier Transform}
\acrodef{edmg}[EDMG]{Enhanced Directional Multi Gigabit}
\acrodef{ekf}[EKF]{Extended Kalman Filter}
\acrodef{elu}[ELU]{Exponential-Linear Unit}
\acrodef{fmcw}[FMCW]{Frequency-Modulated Continuous-Wave}
\acrodef{fov}[FOV]{Field-of-View}
\acrodef{ft}[FT]{Fourier Transform}
\acrodef{fo}[FO]{Frequency Offset}
\acrodef{fpga}[FPGA]{Field Programmable Gate Array}
\acrodef{fr}[FR]{Frequency Range}
\acrodef{frt}[FRT]{Fraction of Resolved Targets}
\acrodef{gpio}[GPIO]{General Purpose Input/Output}
\acrodef{gsps}[GSPS]{Giga-Samples per Second}
\acrodef{gps}[GPS]{Global Positioning Systems}
\acrodef{har}[HAR]{Human Activity Recognition}
\acrodef{ht}[HT]{High Throughput}
\acrodef{if}[IF]{Intermediate Frequency}
\acrodef{ifs}[IFS]{Inter-Frame Spacing}
\acrodef{iaa}[IAA]{Iterative Adaptive Approach}
\acrodef{iht}[IHT]{Iterative Hard Thresholding}
\acrodef{isac}[ISAC]{Integrated Sensing And Communication}
\acrodef{isi}[ISI]{Inter-Symbol Interference}
\acrodef{idft}[IDFT]{Inverse DFT}
\acrodef{jcs}[JCS]{Joint Communication \& Sensing}
\acrodef{nnjpdaf}[NN-JPDAF]{Nearest-Neighbors Joint Probabilistic Data Association Filter}
\acrodef{lo}[LO]{Local Oscillator}
\acrodef{los}[LOS]{Line-of-Sight}
\acrodef{ls}[LS]{Least Squares}
\acrodef{lbm}[LBM]{Loop-Back Memory}
\acrodef{mae}[MAE]{Mean Absolute Error}
\acrodef{mse}[MSE]{Mean Squared Error}
\acrodef{mcs}[MCS]{Modulation and Coding Scheme}
\acrodef{md}[$\mu$D]{micro-Doppler}
\acrodef{mimo}[MIMO]{Multiple Input Multiple Output}
\acrodef{mmwave}[mmWave]{Millimeter Wave}
\acrodef{msps}[MSPS]{Mega-Samples per Second}
\acrodef{mu}[MU]{Multiple User}
\acrodef{music}[MUSIC]{MUltiple SIgnal Classification}
\acrodef{mimo}[MIMO]{Multiple-Input Multiple-Output}
\acrodef{nac}[NAC]{Normalized Auto Correlation}
\acrodef{nco}[NCO]{Numerical Controlled Oscillator}
\acrodef{nlos}[NLOS]{Non-Line-of-Sight}
\acrodef{nls}[NLS]{Non-linear Least-Squares}
\acrodef{ofdm}[OFDM]{Orthogonal Frequency Division Multiplexing}
\acrodef{omp}[OMP]{Orthogonal Matching Pursuit}
\acrodef{per}[PER]{Packet Error Rate}
\acrodef{phy}[PHY]{Physical Layer}
\acrodef{pl}[PL]{Programmable Logic}
\acrodef{pov}[PoV]{Point-of-View}
\acrodef{po}[PO]{Phase Offset}
\acrodef{ps}[PS]{Processing System}
\acrodef{pdp}[PDP]{Power-Delay Profile}
\acrodef{ran}[RAN]{Radio Access Network}
\acrodef{rf}[RF]{Radio Frequency}
\acrodef{rx}[RX]{receiver}
\acrodef{rfsoc}[RFSoC]{Radio Frequency System on a Chip}
\acrodef{rmse}[RMSE]{Root Mean Squared Error}
\acrodef{rss}[RSS]{Received Signal Strength}
\acrodef{rom}[ROM]{Read Only Memories}
\acrodef{rcs}[RCS]{Radar Cross Section}
\acrodef{rpo}[RPO]{Random Phase Offset}
\acrodef{sc}[SC]{Single Carrier}
\acrodef{sdr}[SDR]{Software Defined Radio}
\acrodef{siso}[SISO]{Single Input Single Output}
\acrodef{sls}[SLS]{Sector Level Sweep}
\acrodef{snr}[SNR]{Signal-to-Noise Ratio}
\acrodef{soc}[SoC]{System on a Chip}
\acrodef{sar}[SAR]{Synthetic Aperture Radar}
\acrodef{spb}[SPB]{Signal Processing Blocks}
\acrodef{srrc}[SRRC]{Square-Root-Raised-Cosine}
\acrodef{ssb}[SSB]{Synchronization Signal Block}
\acrodef{ssr}[SSR]{Super Sample Rate}
\acrodef{sta}[STA]{Station}
\acrodef{stf}[STF]{Short Training Field}
\acrodef{stft}[STFT]{Short Time Fourier Transform}
\acrodef{su}[SU]{Single User}
\acrodef{svd}[SVD]{Singular Value Decomposition}
\acrodef{tf}[TF]{Time-Frequency}
\acrodef{toa}[ToA]{Time of Arrival}
\acrodef{tdoa}[TDoA]{Time Difference of Arrival}
\acrodef{to}[TO]{Timing Offset}
\acrodef{tx}[TX]{transmitter}
\acrodef{usrp}[USRP]{Universal Software Radio Peripheral}
\acrodef{ue}[UE]{User Equipment}
\acrodef{uwb}[UWB]{Ultra-Wide Band}
\acrodef{vht}[VHT]{Very High Throughput}
\acrodef{wlan}[WLAN]{Wireless Local Area Network}
\acrodef{wvd}[WVD]{Wigner-Ville Distribution}

{
\theoremstyle{plain}

 }

\copyrightyear{2024}
\acmYear{2024}
\setcopyright{rightsretained}
\acmConference[SENSYS '24]{The 22nd ACM Conference on Embedded Networked Sensor Systems}{November 4--7, 2024}{Hangzhou, China}
\acmBooktitle{The 22nd ACM Conference on Embedded Networked Sensor Systems (SENSYS '24), November 4--7, 2024, Hangzhou, China}\acmDOI{10.1145/3666025.3699357}
\acmISBN{979-8-4007-0697-4/24/11}

\begin{document}
% \fancypagestyle{firststyle}
% {
%     \fancyhf{}
%     \chead{2022 21st ACM/IEEE International Conference on Information Processing in Sensor Networks (IPSN)}
%     \lfoot{\vspace{5pt} 978-1-6654-9624-7/22/\$31.00 ©2022 \\
%     IEEE DOI 10.1109/IPSN54338.2022.0001}
%    \renewcommand{\headrulewidth}{0pt} % removes horizontal header line
% }

\title{HiSAC: High-Resolution Sensing with \\Multiband Communication Signals}

\author{Jacopo Pegoraro}
\email{jacopo.pegoraro@unipd.it} 
\affiliation{
\institution{University of Padova}
\city{Padova}
\country{Italy}}

\author{Jesus O. Lacruz}
\email{jesusomar.lacruz@imdea.org} 
\affiliation{
\institution{IMDEA Networks Institute}
\city{Madrid}
\country{Spain}}

\author{Michele Rossi}
\email{michele.rossi@unipd.it} 
\affiliation{
\institution{University of Padova}
\city{Padova}
\country{Italy}}

\author{Joerg Widmer}
\email{joerg.widmer@imdea.org} 
\affiliation{
\institution{IMDEA Networks Institute}
\city{Madrid}
\country{Spain}}

\settopmatter{printacmref=true, printccs=true, printfolios=false}
% \settopmatter{printacmref=true, printccs=true, printfolios=false}

%Comment the following line for camera-ready
% \renewcommand\footnotetextcopyrightpermission[1]{} % removes footnote with conference information in first column

%-------------------------------------------------------------------------------
\begin{abstract}
\ac{isac} systems are expected to perform accurate radar sensing while having minimal impact on communication. Ideally, sensing should only \textit{reuse} communication resources, especially for spectrum which is contended by many applications. However, this poses a great challenge in that communication systems often operate on narrow subbands with low sensing resolution. Combining contiguous subbands has shown significant resolution gain in active localization. However, multiband \ac{isac} remains unexplored due to communication subbands being highly sparse (\textit{non-contiguous}) and affected by phase offsets that prevent their aggregation (\textit{incoherent}).
To tackle these problems, we design \name{}, the first multiband \ac{isac} system that combines diverse subbands across a wide frequency range to achieve super-resolved passive ranging. 
To solve the non-contiguity and incoherence of subbands, \name{} combines them progressively, exploiting an anchor propagation path between transmitter and receiver in an optimization problem to achieve phase coherence.
\name{} fully reuses pilot signals in communication systems, applies to different frequencies, and can combine diverse technologies, e.g., \acs{5g} and WiGig.
We implement \name{} on an experimental platform in the millimeter-wave unlicensed band and test it on objects and humans. Our results show it enhances the sensing resolution by up to 20 times compared to single-band processing while occupying the same spectrum.
\end{abstract}




\maketitle
% \thispagestyle{firststyle}
%Reset the acronyms 
\acresetall

\section{Introduction}\label{sec:intro}

Endowing wireless communication systems with radar sensing capabilities is one of the key objectives of 3GPP \ac{6g} and future Wi-Fi~\cite{lu2024integrated}. 
In recent years, so-called \ac{isac} systems have enabled a wide range of applications from multitarget tracking~\cite{wu2020mmtrack, pegoraro2023rapid}, person identification~\cite{pegoraro2022sparcs, yang2020mu}, activity recognition~\cite{meneghello2022sharp,lin2023human}, vital signs monitoring~\cite{zeng2020multisense}, pose estimation~\cite{ji2023construct,jiang2020towards}, and object imaging~\cite{zhang2021mmeye, yao2024wiprofile}.

\textbf{Motivation.} A fundamental trade-off in \ac{isac} systems is to achieve high sensing resolution, i.e., the capability of distinguishing multiple closely located targets, and accuracy with minimal impact on the primary communication functionality. 
Ideally, sensing should be performed by fully reusing resources available to the communication system in time and frequency. Particular attention has to be put on spectrum, which is becoming increasingly scarce due to the ubiquitous applications of \ac{rf} transmissions~\cite{testolina2024modeling}. Indeed, \ac{isac} benefits from using a large bandwidth since this is inversely proportional to the ranging resolution, i.e., the minimum signal propagation distance below which two targets can not be distinguished. Note that enhancing ranging resolution is also highly beneficial to advanced sensing applications such as human respiration monitoring. Although small chest displacements can be monitored even with a narrowband signal~\cite{zhang2021exploring}, using the carrier phase, separating the respiration of multiple closely-located subjects with low ranging resolution is extremely challenging and prone to errors

However, the bandwidth available to existing communication systems is insufficient to achieve the desired \textit{cm-level ranging} resolution in \ac{6g}. Even wideband \ac{5g} channels and IEEE 802.11ay (WiGig) in the \ac{mmwave} band can at most reach 37~cm and 17~cm resolution with 400~MHz and 1.76~GHz bandwidth, respectively. Such resolution can be improved by applying super-resolution algorithms based on \ac{music} or compressed sensing, e.g.,~\cite{samczynski20215g, kotaru2015spotfi, khalilsarai2020wifi}, but the bandwidth limitation remains. 

A possible solution to enhance the resolution is to \textit{combine} multiple communication frequency bands to increase the sensing bandwidth. This approach has been attempted for active localization in \ac{ofdm} systems (where the user carries a communication device) \cite{kotaru2015spotfi, xiong2015tonetrack, wan2023multiband, noschese2020multi}, radar \cite{cuomo1999ultrawide, xiong2017coherent,zhang2017multiple}, and recently for sub-6 GHz Wi-Fi sensing in \cite{wang2024uwb}. However, several limitations make the above methods unsuitable for \ac{isac}. 
On the one hand, active localization approaches exploit either contiguous or closely-spaced subbands, which may not be available in \ac{isac} since the spectrum is contended by a plethora of services and contains frequency gaps. Moreover, they can count on a \textit{collaborative} localized device, which simplifies the problem since synchronization errors that prevent accurate delay measurements can be resolved via handshaking \cite{khalilsarai2020wifi}. On the other hand, radar methods use dedicated waveforms, optimized for sensing purposes, and relatively wide subbands to be combined, which significantly simplifies the problem. 
Lastly, \cite{wang2024uwb} employs a neural network model to overcome the above limitations in Wi-Fi, but this ties the system to the specific frequency band, communication technology (Wi-Fi), and hardware used to collect the training data. Conversely, we aim to develop a system that seamlessly adapts to different modulation types such as \ac{5g} \ac{ofdm} and IEEE~802.11ay \ac{sc}.

\textbf{Challenges.} Designing such a system presents several open challenges. First, one must tackle the non-contiguity of communication systems subbands, which may include gaps of several hundreds of MHz or even GHz. Second, the different subbands are affected by time-varying and unknown timing, frequency, and phase offsets that prevent the coherent combination of the \ac{cfr} estimated by the communication protocol over different \ac{isac} \acp{rx}~\cite{zhang2022integration, wu2024sensing}. Although the compensation of timing and frequency offsets in \ac{isac} systems has been widely studied~\cite{zhang2020perceptive,pegoraro2024jump,zhu2018pi,ni2021uplink,li2022csi}, \textit{phase} synchronization is not well investigated since it is not needed in typical sensing tasks such as target tracking and Doppler estimation. On the contrary, achieving phase coherence is a strict requirement to combine multiple subbands over a wide frequency range.
Third, communication subbands are relatively narrow with respect to the total bandwidth required to achieve high resolution. This makes it difficult to model individual subbands as they contain insufficient frequency samples. Conversely, reconstructing the \ac{cfr} over the total bandwidth entails huge computational complexity due to the high number of subcarriers.
Finally, the designed method should generalize to different communication systems (\ac{ofdm} vs. \ac{sc}), protocols (\ac{5g} vs. Wi-Fi), and channel representations (e.g., \ac{cfr} vs. \ac{cir}). 

\begin{figure}
    \centering
    \includegraphics[width=\linewidth]{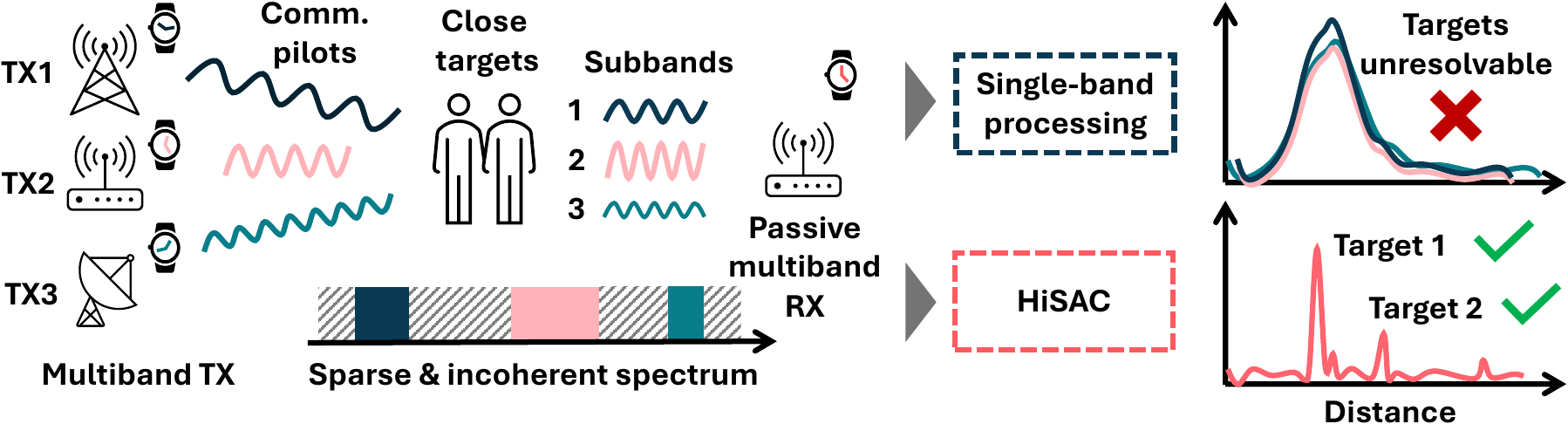}
    \caption{Overview of \name{}'s multiband sensing.}
    \label{fig:concept}
\end{figure}

\textbf{Contribution.} To address these challenges, we design and validate \name{}, the first multiband \ac{isac} system that fully reuses communication traffic across multiple bands (and technologies) to boost the sensing resolution, as shown in \fig{fig:concept}. \name{} first combines all the subbands used by the same \ac{isac} \ac{tx}-\ac{rx} pair (a \textit{subsystem}), which are affected by the same offsets and hence phase-coherent. Then, it compensates for relative timing, frequency, and phase offsets across different subsystems, which are instead incoherent. For this, a new \textit{phase synchronization} algorithm is proposed based on a simple, yet effective, initialization, based on an anchor propagation path, and refinement through an optimization problem. Then, \name{} combines all the available subbands across subsystems with a focused \ac{omp} algorithm~\cite{eldar2012compressed} that exploits the (coarse) prior knowledge about targets' locations obtained from the single subsystems, and outputs super-resolved range estimates. As a final step, \name{} can combine range estimates obtained from different packets or \ac{ofdm} slots \textit{over time} (coherently or incoherently) to further boost the resolution and accuracy.
Our approach combines subbands over several GHz of bandwidth, fully reusing pilot signal in communication systems, e.g., \acp{ssb} in \ac{5g}, and applies to different frequencies, communication systems, and even different technologies, e.g., \ac{sc} and \ac{ofdm}.

We implement \name{} on a \ac{rfsoc} platform in the \ac{mmwave} unlicensed band (58-64~GHz). We demonstrate that \name{} achieves a few-cm ranging resolution on metal and human targets, giving a 3 to 20 times improvement over baseline methods. Moreover, it works in mono-/bi-static configurations on typical multiband systems employing carrier aggregation, bandwidth part, and it is robust to target motion.

The contributions of our work can be summarized as:

1. We propose \name{}, the first \textit{multiband} \ac{isac} system that achieves super-resolution passive ranging using non-contiguous, narrow, and incoherent subbands estimated by sets of communication pilot signals over time.

2. \name{} features new model-based signal processing steps to achieve phase-coherence among subbands that adapt to different systems and technologies across GHz-wide bands.

3. Our approach entails zero additional overhead on communication and seamlessly integrates with multiband communication systems that adopt carrier aggregation or bandwidth part.  

4. We prototype \name{} in the unlicensed \ac{mmwave} band and test it on a vast experimental campaign, showing it can achieve up to 20 times better resolution compared to a single band with the same spectrum occupation per time slot.

\section{Preliminaries and motivation}

In this section, we provide useful background on the applicability of \name{} in \ac{isac} systems, ranging resolution, and phase incoherence due to phase offsets. Then, we highlight the novelty of our approach with respect to existing active multiband systems for localization.

\subsection{\name{} use cases in \ac{isac}}\label{sec:use-cases}

Multiband operation is widely used in communications to increase the data rate and multiplex applications to different users. In this section, we provide an overview of three practical use cases of our method for multiband \ac{isac}: Two of them are typical of cellular networks, such as \ac{5g} or future \ac{6g}, and one tackles cross-technology multiband sensing.

\textbf{Carrier aggregation.} Carrier aggregation is a function implemented in the \ac{ran} and \acp{ue} of \ac{5g} mobile wireless networks. carrier aggregation combines multiple frequency allocations (carriers) at different radio cells to boost the data rate of the connection~\cite{li2023ca}. 
A set of serving cells is identified that contributes to the aggregation, which can take place within the same frequency band (intra-band), e.g., within the sub-$6$ GHz range, or across multiple bands (inter-band), e.g., across \ac{fr}~$2$ and \ac{fr}~$3$.
Exploiting the channel estimation process carried out on each frequency band for sensing would greatly enhance the available bandwidth and, as a result, the ranging resolution.

\textbf{Bandwidth part.} Bandwidth part is a mechanism to split the radio channel available to a cell into multiple segments (\textit{parts}), which can then be used to allocate different regions of the spectrum to different applications (or \acp{ue})~\cite{lin2021primer}. Only one part can be active at a time, and \acp{ssb} can be transmitted in each part to synchronize to different  \acp{ue}. Given that \acp{ssb} have a relatively narrow bandwidth, it is appealing to develop a system that can combine the \acp{ssb} transmitted by one or more radio cells to perform accurate mono-static ranging by exploiting the total frequency aperture over a wider bandwidth. Note that, bandwidth part poses the additional requirement that the system must be able to operate with \ac{cfr} estimates that are not collected simultaneously in all the subbands.

\textbf{Cross-technology multiband processing.} Coexistence of \textit{unlicensed} \ac{5g} and IEEE 802.11ay has been advocated in the 60 GHz band and has gathered significant interest from academia and industry \cite{daraseliya2021coexistence, qualcomm}. Flexible multiband processing across \ac{ofdm} and \ac{sc} communication technologies is appealing to boost sensing resolution in these cases.  While combining multiple Wi-Fi channels has been studied for both communication~\cite{barrachina2021wi} and localization~\cite{kotaru2015spotfi, xiong2015tonetrack}, combining multiple frequency bands obtained by Wi-Fi and cellular communication systems is still unexplored.  
In the \ac{mmwave} band, IEEE~802.11ay channels span a wide bandwidth of $1.76$~GHz, while in \ac{5g} channels are limited to $400$~MHz. Therefore, combining multiple channels across the two technologies can yield practical \ac{uwb}-level resolution with \ac{cfr} estimates spanning several GHz of bandwidth.

\subsection{Delay and ranging resolution}\label{sec:delay-resolution}

The delay resolution, $\Delta \tau$, of a localization or passive sensing system is related to the bandwidth, $B$, of the transmission signal as $\Delta \tau = 1/B$. The corresponding ranging resolution for passive sensing also depends on the angle between the segments connecting the \ac{tx} to the target and the target to the \ac{rx} (\textit{bi-static} angle), $\beta$, as $\Delta r = c /[2B \cos\left(\beta/2\right)]$.
A mono-static system, with co-located \ac{tx} and \ac{rx}, gives $\Delta r = c/(2B)$ which minimizes $\Delta r$ with respect to $\beta$.
Unlike radar systems, which typically feature a large transmission bandwidth fully dedicated to sensing, communication systems are relatively narrowband for sensing purposes. As an example, even \ac{mmwave} \ac{5g} system with $400$ MHz channels can only reach up to $37$ cm mono-static ranging resolution, which may be insufficient for fine-grained sensing applications. Moreover, such resolution is only obtained if the full bandwidth is used to estimate the channel. This is often not the case, since pilot signals are transmitted on a subset of the available subcarriers. The \acp{ssb} used for synchronization with \acp{ue}, for example, occupy $240$ \ac{ofdm} subcarriers with at most $240$~kHz subcarrier spacing. This leads to a very coarse ranging resolution of $\Delta r = c/(2\cdot 240 \cdot 240 \mathrm{kHz}) = 2.6$~m. 

Improving ranging resolution is a challenging task since the bi-static angle depends on the location of the \ac{tx}, \ac{rx}, and target, and increasing the bandwidth is not viable in \ac{isac} systems since it is pre-determined by the communication protocol. Super-resolution approaches have been proposed that exploit subspace-based methods, e.g., \ac{music}~\cite{kotaru2015spotfi}, or compressed sensing algorithms~\cite{khalilsarai2019wifi}, exploiting assumptions on the structure of the channel (e.g., sparsity). However, none of these methods can drastically improve the ranging resolution, and limitations due to the narrow bandwidth remain.

\subsection{Phase-incoherence among subbands}

In a wireless communication system, the clock signal of each node is generated from a \ac{lo}. The different \acp{lo} are asynchronous, meaning that, due to hardware non-idealities, they are subject to time-varying relative drifts from their nominal oscillating frequencies~\cite{zhang2021overview}. In addition, their initial phase is random.
This introduces unwanted offsets in the received signals that are specific to each \ac{tx}-\ac{rx} pair. These can be categorized into \ac{to}, \ac{cfo}, and \ac{rpo}~\cite{wu2024sensing}. 

\ac{to} results from the lack of time synchronization between the \ac{tx} and \ac{rx}. It is due to the unknown shift or offset affecting the \ac{rx} clock relative to the \ac{tx} one, and to the synchronization point chosen by the \ac{rx}. \ac{to} is time-varying and causes an undesired phase term that increases linearly with the subcarriers in \ac{ofdm} systems.

\ac{cfo} is due to the time-varying difference in the \acp{lo} of the \ac{tx} and \ac{rx}. Communication systems typically estimate and partially compensate for the \ac{cfo}. This leads to a residual \ac{cfo} that is fast time-varying, as a result of the compensation error~\cite{wu2024sensing}. \ac{cfo} causes a cumulative phase shift across packets or \ac{ofdm} slots.  
	
The \ac{rpo} can be caused by non-idealities in the \ac{tx} and \ac{rx} devices, as well as by phase noise~\cite{rasekh2019phase}. It varies on an \ac{ofdm} symbol basis. Note that \ac{rpo} can be present even between the multiple channels of the same \ac{lo}.

When multiple \ac{isac} systems in different frequency bands are considered, a direct combination of their \acp{cfr} is infeasible due to the presence of the above offsets. Indeed, this causes the phases of the \acp{cfr} in the different bands to be misaligned at the different \ac{rx}s, preventing the construction of a common model spanning the full frequency band. 
Several approaches have been proposed to tackle \ac{to} and \ac{cfo} in \ac{isac} systems~\cite{zhang2020perceptive,pegoraro2024jump,zhu2018pi,ni2021uplink,zeng2019farsense,zeng2020multisense,li2022csi,chen2023kalman,meneghello2022sharp}. However, none of these tackles phase synchronization by also eliminating the \ac{rpo}, which is essential for multiband \ac{cfr} combination.

\subsection{Innovation over multiband localization}\label{sec:innovation}

Exploiting multiple frequency bands for localizing an \textit{active} receiver device has been widely studied. \name{} instead tackles the different problem of \textit{passive} ranging, in which the target of the localization is an object or person, not necessarily equipped with a radio device. 

Two main aspects prevent the application of existing methods for active localization to the passive \ac{isac} scenario. 

First, active localization targets the time-of-arrival estimation of the \ac{los} path between TX and RX, as done by Chronos~\cite{vasisht2016decimeter} and SpotFi~\cite{kotaru2015spotfi}, among others~\cite{bansal2021owll, khalilsarai2020wifi}. Hence, multipath reflections are regarded as a nuisance and have to be compensated for. Conversely, our work focuses on obtaining the propagation delays of reflections on passive objects that are causing the multipath effect. These are significantly weaker than the \ac{los}, due to the longer propagation distance and scattering on the target. The applicability of active localization methods for paths other than the \ac{los} has not been demonstrated, making them unsuitable for radar-like sensing, which is the objective of \name{}.  

Second, existing multiband localization methods combine overlapping, contiguous, or relatively close frequency subbands~\cite{kotaru2015spotfi, xie2015precise, xiong2015tonetrack}. This is due to the limitations of existing algorithms to achieve phase coherence, which do not scale well to wide frequency bands, and the high computational complexity of compressed sensing. Due to these limitations, existing methods can not be applied to \ac{isac}, where subbands may be widely separated in frequency as discussed in \secref{sec:use-cases}, with a low ratio of measured frequencies over the total bandwidth aperture. We demonstrate instead that \name{} can accurately perform ranging with as low as $1/6$ measured frequencies ratio (see \secref{sec:indepth-eval}). This is enabled by exploiting the channel sparsity, originally applying compressed sensing in two incremental steps to reduce complexity, and to the increased robustness of \name{}'s phase offsets compensation method. Thanks to these innovations, \name{} is also the first multiband system that demonstrates \textit{cross-technology} sensing capabilities, combining \ac{ofdm} and \ac{sc} channel estimates over wide bands.

Third, active localization relies on \textit{information exchange} among nodes to compensate for phase offsets, which introduces overhead on communication and energy consumption. Multiband methods like Chronos~\cite{vasisht2016decimeter}, Owll~\cite{bansal2021owll}, M$^3$~\cite{chen2019m3}, and others~\cite{khalilsarai2019wifi, khalilsarai2020wifi} exchange channel measurements using dedicated handshaking protocols and perform active frequency hopping. The overhead due to these additional transmissions reduces the Wi-Fi throughput by 18.5$\%$ in Chronos~\cite{vasisht2016decimeter}, while \cite{bansal2021owll} shows their significant impact on the battery life of embedded devices. 
Conversely, \name{} does not require exchanging channel estimates or performing frequency hopping. It only uses channel estimates that are \textit{already available} at the receiver, without introducing any further cooperation or information exchange (i.e., avoiding the involved communication overhead).

\section{System model}
\label{sec:cfr-model}

In this section, we formulate a general model of a multiband \ac{isac} system that fits all the use cases in \secref{sec:use-cases}.

\subsection{Non-coherent subsystems and subbands}

\begin{figure}[t!]
    \centering
    \includegraphics[width=0.9\linewidth]{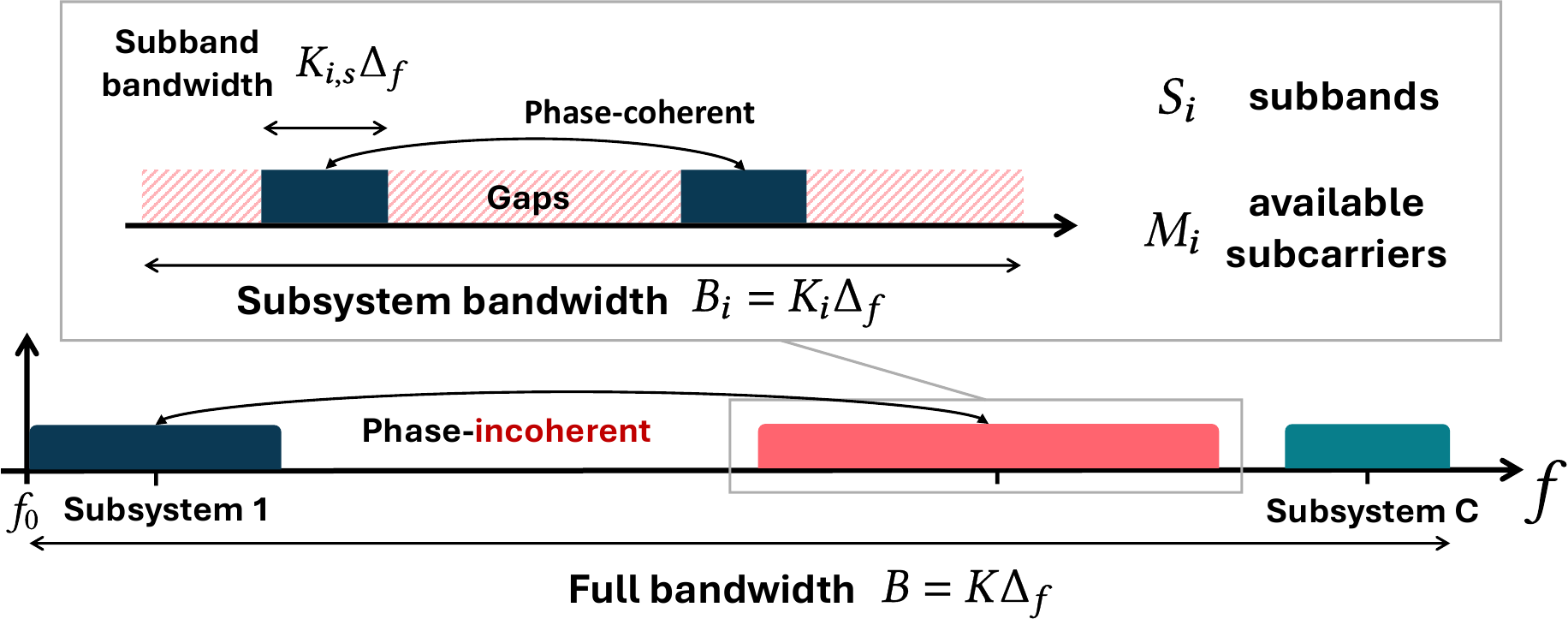}
    \caption{Summary of subsystems and subbands notation.}
    \label{fig:subbands}
\end{figure}

Consider a wide frequency band with bandwidth $B$ and starting frequency $f_0$, denoted as the \textit{full band} of interest, as shown in \fig{fig:subbands}. 
The full band represents the total frequency aperture of \name{}, which gives the ideal delay resolution of the system. In practice, the full band may represent the total aperture of carrier aggregation or bandwidth part systems, spanning from the frequency of the first estimated subcarrier in the channel response to the last one, including frequency gaps. 

We call $\Delta_f$ the spacing of the frequency samples of the considered \ac{cfr} (subcarriers). $\Delta_f$ corresponds to the subcarrier spacing in \ac{ofdm} systems or to the \ac{dft} samples spacing used in \ac{sc} systems. The total number of \textit{virtual} subcarriers in the full band is $K=B/\Delta_f$, indexed by $k=0, \dots, K-1$. We use the term virtual to highlight that not all the subcarriers are used for communication, which is carried out on a subset of the full band spectrum. In the following, we assume that all the considered subbands share the same subcarrier spacing $\Delta_f$, which can be achieved using interpolation or downsampling.

We consider an \ac{isac} system consisting of $C$ non-coherent subsystems affected by \ac{to}, \ac{cfo}, and \ac{rpo}. Each subsystem, $i$, includes one \ac{tx}-\ac{rx} pair that may be co-located (\textit{mono-static}) or widely separated (\textit{bi-static}). In practice, a subsystem can be represented by co-located \acp{bs} or \acp{ap} from different operators, acting as mono-static \ac{isac} transceivers, or \ac{bs}-\ac{bs}/\ac{bs}-\ac{ue} pairs in the bi-static case. Subsystem $i$ has bandwidth $B_i$, starting from frequency $f_i$, contained in the full band. The total number of virtual subcarriers of a subsystem is $K_i = B_i/\Delta_f$.
Within each subsystem, the channel is estimated over a set $\mathcal{S}_i$ of potentially non-contiguous subbands, with $|\mathcal{S}_i|=S_i$. The subbands may span the whole $B_i$ or a part of it, according to the allocation of pilot signals used for channel estimation. We use index $s = 1, \dots, S_i$ to identify the subbands in subsystem $i$. Note that $S_i$ may equal $1$ if system $i$ has a single subband. 
We call the total number of subbands in the system $S=\sum_{i=1}^C S_i$. Each subband contains a set of $K_{i, s}$ subcarriers. The subcarriers in the system for which the channel is estimated are called \textit{available} subcarriers. The number of available subcarriers is $M_i = \sum_{s=1}^{S_i} K_{i, s}$, for subsystem $i$ and 
$M = \sum_{i=1}^C M_i$ for the whole system, with $M_i \leq K_i$ and $M < K$. 

Note that \name{} handles subsystems and subbands with different bandwidths, which is critical to seamlessly integrate it into communication systems operating in diverse frequency bands.

Since each \ac{isac} subsystem has a single \ac{rx}, all the subbands of the same subsystem $i$ share the same \ac{to}, \ac{cfo}, and \ac{rpo} due to being implemented on the same radio device. 
In our model, we consider \ac{to}, \ac{cfo}, and \ac{rpo} of subsystem $i$ to be relative to the first subsystem ($i=1$), which we take as a reference. Hence, we denote by $\tau_{o, i}(t)$, $f_{o, i}(t)$, and $\varphi_{o, i}(t)$ the relative \ac{to}, \ac{cfo}, and \ac{rpo} of subsystem $i$, respectively. The absolute offsets do not impact the performance of our system and are omitted in the model.

\subsection{Multiband channel model}

In this section, we present the multiband \ac{cfr} model.
We consider a time-varying multipath channel with $L$ propagation paths, where $t$ is used to denote time. We denote by $\tau_l(t)$ and $\alpha_l(t)$ the delay of the $l$-th channel path due to propagation and its complex amplitude at time $t$, accounting for the combined effect of the propagation loss and the target's scattering phase~\cite{richards2010principles}.
\tab{tab:notation} summarizes the notation used in the system model.

We model the \ac{cfr} over the full band, discretized by the subcarrier spacing $\Delta_f$. 
The expression of the full band \ac{cfr} for subcarrier $k$ at time $t$ is
\begin{equation}\label{eq:fullband-cfr}
    H_k(t) = \sum_{l = 1}^{L} \alpha_{l}(t) e^{ -j 2\pi k\Delta_f \tau_l(t)}, \quad k=0, \dots, K-1,
\end{equation}
with delay resolution $\Delta\tau = 1/(\Delta_f K)$.
The \ac{cir} can be obtained from the \ac{cfr} via an \ac{idft} along the subcarriers.
Note that, in \eq{eq:fullband-cfr}, we include the carrier phase into each path's complex amplitude $\alpha_l(t)$. Considering the carrier phase in the full band channel model would lead to high sensitivity of the algorithm to errors in the positioning of the \ac{tx} and \ac{rx} antennas of each subsystem, which would make it impractical to use. This is especially true for at \ac{mmwave} frequencies where the wavelength is short. Hence, \name{} exploits the bandwidth aperture $B$, rather than the carrier $f_0$.  
As commonly done in the \ac{uwb} channels literature~\cite{wan2024ofdm, wan2022fundamental}, we consider the coefficients $\alpha_l(t)$ to be constant within the frequency band of interest. This holds if the total bandwidth is less than 20\% of the carrier frequency~\cite{molisch2009ultra}.

We denote by $k_{i, s}$ the starting index of subband $s$ in subsystem~$i$ in the full band grid, $k=0,\dots, K$.
The \ac{cfr} in subband $s$, at subcarrier $\kappa = 0, \dots, K_{i,s}-1$, is
\begin{equation}\label{eq:subband-cfr-carrier}
    H_{i, s, \kappa}(t) = e^{j\phi_{o,i}(t)}e^{-j2\pi \kappa\Delta_f \tau_{o,i}(t)}\sum_{l = 1}^{L} \alpha_{l}(t) e^{ -j 2\pi (k_{i,s} + \kappa)\Delta_f\tau_l(t)},
\end{equation}
where $\phi_{o, i}(t) = -2\pi f_{o,i}(t)t + \varphi_{o,i}(t)$ is denoted by \ac{po} in the following. The \ac{po} contains the contribution of the \ac{cfo} and the \ac{rpo} since these are constant in $\kappa$ and $l$. 

\begin{table} [t!]
\small
	\begin{center}
		\begin{tabular}{lrlr}
\toprule
Start frequency & $f_0$ &No. avail. subc. subsys. $i$ & $M_i$\\
Full bandwidth& $B$ & \ac{to} subsystem $i$ & $\tau_{o, i}$\\
         Subcarrier spacing      & $\Delta_f$ & \ac{cfo} subsystem $i$  &$f_{o, i}$ \\
            Total no. subcarriers  & $K$ & \ac{rpo} subsystem $i$  & $\varphi_{o, i}$\\
             No. of subsystems  & $C$ & \ac{po} subsystem $i$ &$\phi_{o, i}$\\
             No. subcarriers subsys. $i$  & $K_i$ &  Total delay resolution&$\Delta\tau$\\
            No. subc. subsys. $i$ subb. $s$  & $K_{i,s}$ & Number of paths &$L$\\
             Start frequency subsys. $i$  & $f_i$ &  Complex gain path $l$&$\alpha_l$\\
              Bandwidth subsystem $i$ & $B_i$ &  Delay path $l$ &$\tau_l$\\
            Subbands set subsystem $i$ & $\mathcal{S}_i$ & Start idx. subsys. $i$ subb. $s$ &$k_{i, s}$\\
             No. subbands subsystem $i$ & $S_i$ &  $k$-th subc. in total \ac{cfr} &$H_{k}$ \\
             No. available subcarriers & $M$ & $k$-th subc. subsys. $i$ subb. $s$ &$H_{i,s,k}$\\
 \bottomrule
\end{tabular}
	\end{center}
	\caption{Summary of the notation.} 
	\label{tab:notation}
\end{table}

\section{\name{}  methodology}\label{sec:method}

\begin{figure}[t!]
    \centering
\includegraphics[width=\linewidth]{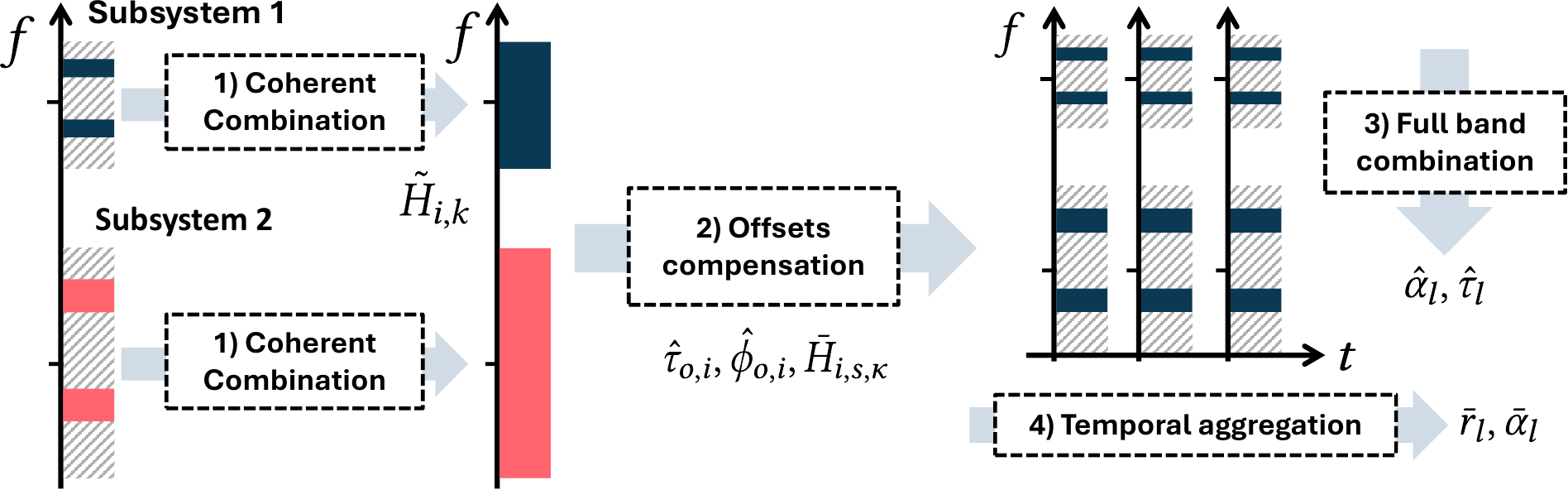}
    \caption{\name{} high-level overview.}
    \label{fig:block-diagram}
\end{figure}

This section presents \name{}'s processing steps, which are summarized in the following and shown in \fig{fig:block-diagram}.

\noindent\textbf{(1) Coherent intra-subsystem combination.} The first step performs a coarse multiband reconstruction of the \ac{cfr}, using only the coherent subbands in each subsystem, as detailed in \secref{sec:intra-subsys}. The reason to first aggregate subbands over each subsystem is to obtain a wider-band \ac{cfr}, to simplify the subsequent removal of phase offsets. 

\noindent\textbf{(2) \ac{to} and \ac{po} compensation.} This step applies a new algorithm to achieve phase synchronization across multiple subsystems, making them suitable for coherent multiband combination (see \secref{sec:po-compensation}). Compared to existing approaches, ours is more robust with narrow, non-contiguous subbands by (i) exploiting an anchor path for \ac{to} initialization and (ii) accurately estimating \ac{to} and \ac{po} via optimization.

\noindent\textbf{(3) Inter-subsystem multiband reconstruction.} Delays and amplitudes of the multipath components in the \ac{cfr} are estimated using all the available subbands. This is done with the \ac{omp} algorithm, to tackle the gaps in the \ac{cfr} measurements, by restricting the search space around the initial estimates obtained from the coherent subsystems to counter the discretization error, as detailed in \secref{sec:omp}. 

\noindent\textbf{(4) Temporal aggregation.} \name{} can optionally aggregate the estimates of the multipath parameters across time, represented by different packets or \ac{ofdm} slots. The aggregation consists of an accumulation and selection algorithm, that yields significantly improved ranging accuracy and resolution after a few time slots (see \secref{sec:aggregation}).

Steps (1)-(3) do not depend on the time instant in which the \ac{cfr} is estimated, so we omit the time $t$ in their description.

\subsection{Intra-subsystem coherent combination}\label{sec:intra-subsys}

As a first step, we combine the subbands obtained by each subsystem $i$ coherently (since they experience the same \ac{to} and \ac{po}) to coarsely estimate the multipath delays and complex amplitudes.
To do so, we first observe that the \ac{cfr} in \eq{eq:fullband-cfr} is \textit{sparse} in the delay domain. This stems from the fact that in typical communication channels the number of paths, $L$, is much lower than the number of total subcarriers, $K$. Hence, it is possible to represent the \ac{cfr} as a combination of much fewer basis signals compared to the number of subcarriers. This fact can be exploited to recover the \ac{cfr} for subsystem $i$ from the incomplete set of $M_i$ \ac{cfr} measurements, according to the compressed sensing principle~\cite{eldar2012compressed}.  

\textbf{Notation and definitions.} To formulate the intra-subsystem \ac{cfr} reconstruction as a compressed sensing problem, we set up a grid with $Q_i$ channel path candidates as
\mbox{$0, \dots, (Q_i-1)\delta_i$},
where $\delta_{\rm i}$ is the grid granularity of subsystem $i$. The size of the grid should be selected according to a trade-off between reconstruction performance and computational complexity. Each candidate is a complex sinusoidal signal that represents a possible channel path with its corresponding delay. To compactly represent the set of candidate paths, we construct a partial Fourier matrix, $\mathbf{F}_i$, that spans all the subcarriers in subsystem $i$ and the delays in the grid. Element $m, q$ of $\mathbf{F}_i$ is $\left( \mathbf{F}_i\right)_{m, q} = e^{j2\pi m q \delta_i \Delta_f} / \sqrt{K_i}$. The columns of $\mathbf{F}_i$ represent the different complex sinusoids corresponding to the candidate paths. However, in our multiband \ac{isac} system not all the subcarriers are observed by subsystem $i$. To model this aspect, we define $\mathbf{A}_i$ as the matrix whose rows are the vectors of all zeros but the $k$-th component, which equals $1$, with $k \in \{k_{i,1}, \dots, K_{i,1}-1, \dots, k_{i,S}, \dots, K_{i,S}-1\}$. We use $\mathbf{A}_i$ to select the rows of $\mathbf{F}_i$ whose indices are in the set of available \ac{cfr} samples in subsystem $i$. We obtain the final model matrix, which only contains the available parts of the complex sinusoids corresponding to the candidate paths, as $\mathbf{\Gamma}_i = \mathbf{A}_i\mathbf{F}_i$, of dimension $M_i\times K_i$. Then, we collect all the available \ac{cfr} measurements in subsystem $i$ into vector $\mathbf{H}_i =[H_{i, 1, 0}, \dots, H_{i, 1, K_{i, 1} - 1}, \dots, H_{i, S, 0}, \dots, H_{i, S, K_{i, S}-1}]^{\mathsf{T}}$, of dimension $M_i$.

\textbf{Compressed sensing problem formulation.} To represent the coarse \ac{cir} estimate obtained by fusing the $S_i$ subbands in subsystem $i$, we use vector $\mathbf{h}_i$, of dimension $K_i$. $\mathbf{h}_i$ is related to the \ac{cfr} in the $i$-th subsystem by the following model $\mathbf{H}_i = \mathbf{\Gamma}_i \mathbf{h}_i + \mathbf{w}_i$, where $\mathbf{w}_i$ is a complex Gaussian noise vector with element-wise variance equal to $\sigma^2_w$.
Since the \ac{cfr} is sparse in the delay domain, only a small fraction of the components of $\mathbf{h}_i$ are non-zero.

Estimating the non-zero components of $\mathbf{h}_i$ is the aim of the compressed sensing-based \ac{cfr} reconstruction. Specifically, this can be done by solving the following optimization problem
\begin{equation}\label{eq:subsystem-cir}
    \mathbf{h}_i = \argmin_{\mathbf{h}} ||\mathbf{h}||_0 \mbox{  subject to  } ||\mathbf{H}_i - \mathbf{\Gamma}_i\mathbf{h}||_2^2 < \epsilon,
\end{equation}
where $||\cdot||_0 $ is the number of non-zero components of a vector and $\epsilon$ is a pre-defined reconstruction error threshold that depends on the noise level. 

\textbf{\ac{omp}-based solution.} To solve the problem in \eq{eq:subsystem-cir}, we use the \ac{omp} algorithm~\cite{eldar2012compressed}, which
operates by iteratively adding non-zero components to $\mathbf{h}_i$ in a greedy fashion. At each iteration, the candidate in $\mathbf{\Gamma}_i$ that best correlates with the residual measurements is selected as valid, and its complex amplitude is estimated via least-squares. The new valid candidate is then subtracted from the measurements and the process is repeated. 

\ac{omp} finds an approximation to the sparsest estimate of the \ac{cir} that leads to a bounded \ac{mse} with the \ac{cfr} measurements. The bound on the \ac{mse} is regulated by the positive constant $\epsilon$, which can be estimated from the \ac{cfr} estimates data. In our implementation, we stop the execution of \ac{omp} once the reconstruction error on the original measurements, $\mathbf{H}_i$, falls below 5\% of the norm of $\mathbf{H}_i$. 

We call $L_i^{\rm OMP}$ the number of non-zero components of $\mathbf{h}_i$ as reconstructed by \ac{omp}.
Together with the complex coefficients of the valid candidate paths, \ac{omp} yields the corresponding path delays. We group the coefficients and delays into the set $\{\alpha_{i, l}, \tau_{i,l}\}$, $l=1, \dots, L_i^{\rm OMP}$. These correspond to the values and locations of the non-zero elements of vector $\mathbf{h}_i$ in the grid of candidates. Note that, since no phase offset compensation has been applied yet, 
$\tau_{i,l}$ is an estimate of $\tau_{1,l} + \tau_{o, i}$, i.e., the delays $\tau_{i,l}$ \textit{contain the relative \ac{to}}. Finally, a synthetic \ac{cfr} for subsystem $i$, is obtained as $\tilde{H}_{i,k} = \sum_{l = 1}^{L_i^{\rm OMP}} \alpha_{i, l} e^{ -j 2\pi k\Delta_f \tau_{i,l}},$
where $k$ can be extended even outside of the subsystems' bandwidth. This will be used in the next section to compensate for relative \acp{to} and \acp{po}.

\subsection{Relative TO and PO compensation}\label{sec:po-compensation}

In the second step, the relative \acp{to} and \acp{po} among the different \ac{isac} subsystems are compensated for. 
This is done by leveraging an \textit{anchor} propagation path between the \ac{tx} and the \ac{rx} of each subsystem since the \ac{to} and \ac{po} are common to all paths~\cite{zhang2021overview}.
The anchor path could be either the \ac{los} path, which is commonly assumed to be available in \ac{isac}~\cite{zhang2022integration}, or a non-\ac{los} static path seen by all subsystems.
In the following, we first cast the \ac{to} and \ac{po} estimation as an optimization problem. Then, we detail how to exploit the anchor path in different subsystems to initialize the \ac{to} accurately. Finally, we solve the optimization and compensate for \ac{to} and \ac{po} to achieve phase-coherence.

\textbf{Problem formulation.} We compensate for $\tau_{o, i}$ and $\phi_{o, i}$ using the synthesized \ac{cfr} of subsystem $i$ and that of the reference subsystem. 
Intuitively, compensating for \ac{to} and \ac{po} amounts to performing the following two operations: (i)~\textit{phase-rotating} the complex values of the \ac{cfr} of subsystem $i$, $\tilde{H}_{i,k}$, by $-\phi_{o, i}$ for all subcarriers, and (ii)~\textit{re-modulating} the \ac{cfr} by multiplying it by the complex exponential $e^{j2\pi k\Delta_f\tau_{o, i}}$. It can be seen that applying (i) and (ii) cancels out the offsets from the \ac{cfr} in \eq{eq:subband-cfr-carrier}.

As a result, \ac{to} and \ac{po} can be estimated by solving the following minimization problem
\small
\begin{eqnarray}\label{eq:nls-offset}
    \{\hat{\tau}_{o,i},
\hat{\phi}_{o,i}\} &= &\argmin_{\tau, \phi} \sum_{k = 0}^{K-1}\left|\tilde{H}_{1, k} - e^{-j\phi}e^{j2\pi k\Delta_f\tau }\tilde{H}_{i,k}\right|^2 \nonumber \\
     & = &\argmin_{\tau, \phi}  \sum_{k = 0}^{K-1} -2 \mathrm{Re} \left\{ e^{-j\phi}e^{j2\pi k\Delta_f\tau }\tilde{H}_{1, k}^{*} \tilde{H}_{i,k}\right\},
\end{eqnarray}
\normalsize
where $^{*}$ and $\mathrm{Re}\{\cdot\}$ are the complex conjugate and real part of a complex number, respectively.
Solving \eq{eq:nls-offset} gives the phase rotation and re-modulation delay that best match the measured \ac{cfr} on the reference subsystem.
The problem is high-dimensional, non-linear, and non-convex, which causes solvers to converge to inaccurate solutions and get stuck in local minima.
Moreover, its computational complexity is prohibitive since the number of frequency samples $K$ is huge due to its relation with the subcarrier spacing. 

\begin{figure}[t!]
	\begin{center}   
		\centering
		\subcaptionbox{Delay difference of the anchor paths initializes the \ac{to}. \label{fig:delays}}[0.45\columnwidth]{\includegraphics[width=0.45\columnwidth]{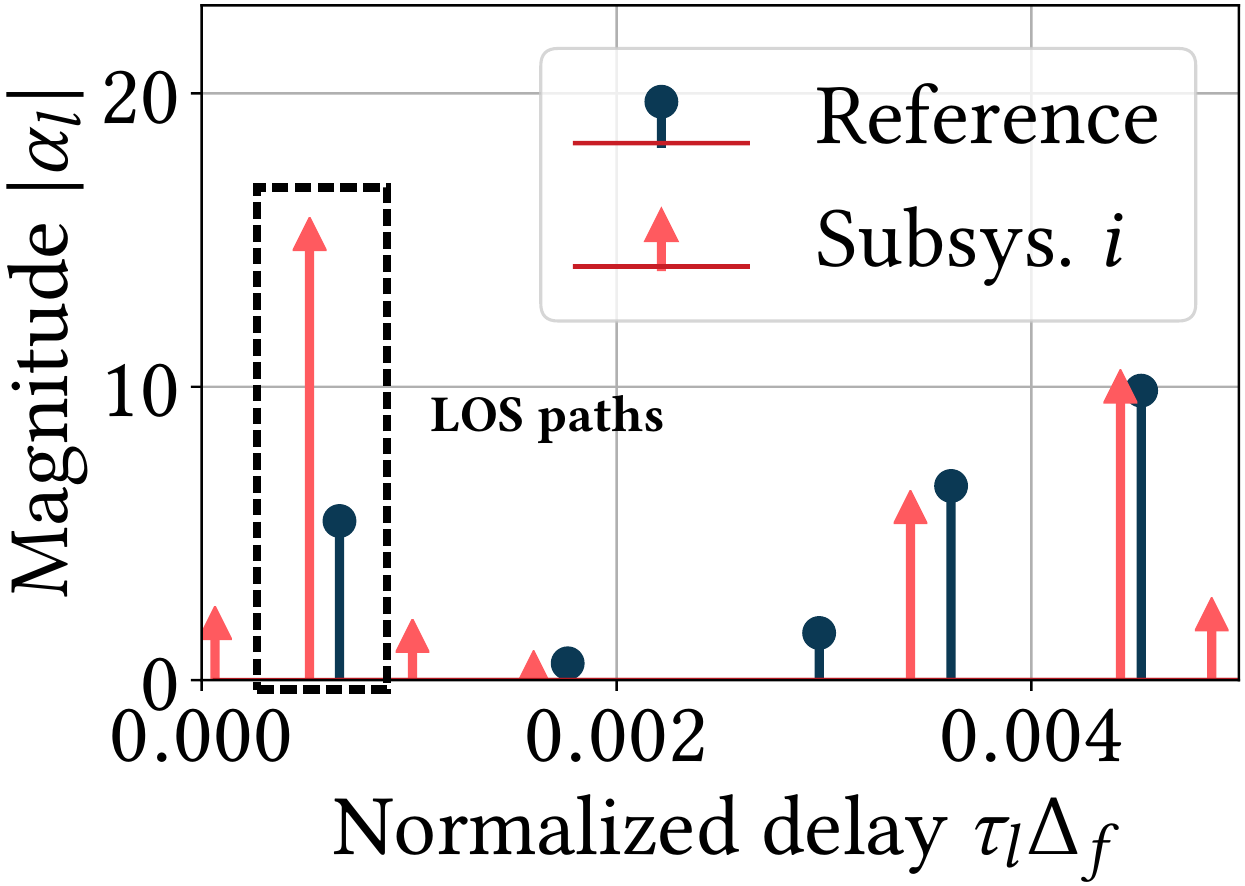}}
		\subcaptionbox{Cost function, yellow/blue represent high/low values. \label{fig:costfun}}[0.45\columnwidth]{\includegraphics[width=0.45\columnwidth]{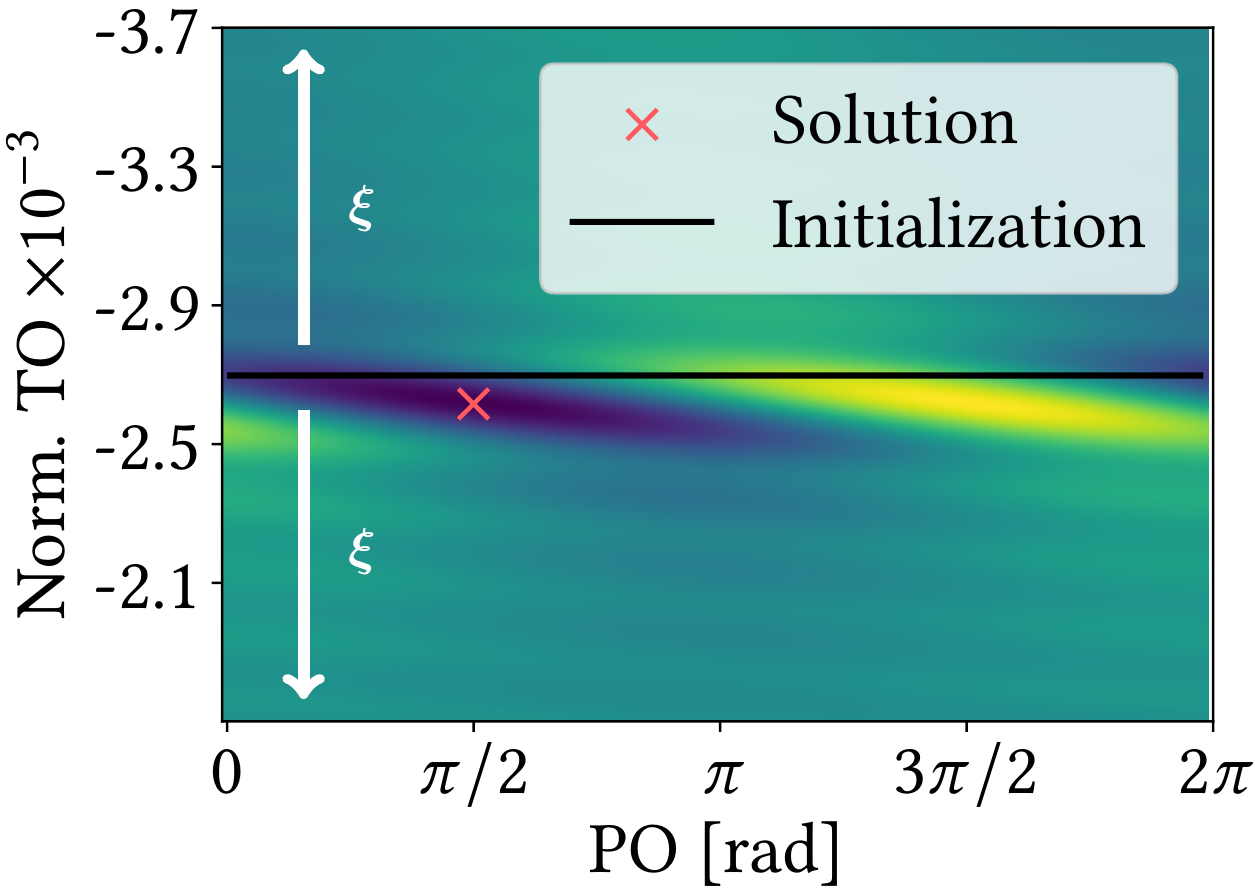}}
		\caption{Example estimation of the initialization value for the \ac{to}, (a), and the refinement via optimization, (b).}
		\label{fig:coherence-examples}
	\end{center}

\end{figure}

\textbf{Initialization.} To solve the problem, we obtain an accurate initial estimate of the \ac{to} from the synthesized \acp{cfr}. We select the anchor path delay for subsystem $i$, denoted by $\tau_{i, 1}$, among $\{\alpha_{i, l}, \tau_{i,l}\}, l=1, \dots, L_i^{\rm OMP}$. If this corresponds to the \ac{los}, it is easily identifiable by having strong received power compared to scattered paths \cite{kuschel2019tutorial} and by having the smallest propagation delay. If the anchor path is a non-\ac{los} static reflection, it can be localized by each subsystem before applying \name{}.
The path delays of the reference subsystem, $\tau_{1, 1}$, are not affected by relative \ac{to}. Since the \ac{to} is common to \textit{all} the propagation paths, it can be estimated from the anchor path and used to initialize the solution to \eq{eq:nls-offset}. We estimate the relative \ac{to} as the difference between the anchor path delay of subsystem $i$ and that of the reference subsystem, $ \hat{\tau}'_{o,i} = \tau_{1, i} - \tau_{1, 1}$. This reasoning is shown in \fig{fig:delays}. The \ac{omp} grid step used in subsystem $i$, $\delta_i$, limits the accuracy of $\hat{\tau}'_{o,i}$, since the values of $\tau_{1, i}$ can only lie on the grid. Therefore, we apply a refinement step to the \ac{to} estimate and also obtain the \ac{po}.

Initializing the \ac{to} estimate using delay differences is an innovation of \name{}. It allows reducing the complexity of directly solving \eq{eq:nls-offset} by reducing the search space for the \ac{to} which, unlike the \ac{po}, is unbounded and causes fast oscillations of the cost function.
Moreover, note that solving \eq{eq:nls-offset} requires achieving phase coherence between the recostructed \acp{cfr} across the \textit{full} bandwidth. When the frequency bands of the subsystems are widely separated, this is challenging since even a small error on the \ac{to} can cause large phase variations across wide frequency bands. Existing methods based on linear fitting of the phase, e.g.,~\cite{kotaru2015spotfi}, are prone to errors in these cases as we show in \secref{sec:results}.

\textbf{Refinement.} Once the initial estimate of the \ac{to} has been obtained, we refine it by solving \eq{eq:nls-offset}. The computational complexity is greatly reduced by searching over a small neighborhood of $\hat{\tau}'_{o, i}$, while for the \ac{po} we search over the interval $[0, 2\pi]$. We use a grid search for this optimization, focusing the search space in $[\hat{\tau}'_{o, i}- \xi, \hat{\tau}'_{o, i}+\xi]$ for the \ac{to}, where $\xi = 5\delta_i$, i.e., 5 times the \ac{omp} grid step of the subsystem. Note that this choice achieves a trade-off between reducing the search space as much as possible and accounting for errors in the initialization, whose accuracy depends on $\delta_i$. We use a grid spacing of $2\xi / 100$ for \ac{to} and $2\pi / 100$ for \ac{po}, thus searching over a $100 \times 100$ grid. The size of the grid can be reduced or increased depending on computational resources.
As a result, we obtain estimates of the \ac{to}, $\hat{\tau}_{o,i}$, and \ac{po}, $\hat{\phi}_{o,i}$.
\fig{fig:costfun} shows an example of the refinement step. The initialization of the \ac{to} allows restricting the search to a small area in the cost function that presents a clear minimum.

\textbf{\ac{to} and \ac{po} compensation.} The \ac{to} and \ac{po} are compensated for in each subband by applying the phase-rotation and re-modulation steps (see points (i) and (ii) in the problem formulation) using the estimates $\hat{\tau}_{o,i}$, and $\hat{\phi}_{o,i}$. This is mathematically expressed as
\begin{equation}
    \bar{H}_{i, s, \kappa} = e^{-j \hat{\phi}_{o, i}}e^{j2\pi \hat{\tau}_{o,i} \kappa \Delta_f}H_{i, s, \kappa},
\end{equation}
where the resulting \ac{cfr} for subband $s$ of subsystem $i$, $\bar{H}_{i, s, \kappa}$, is now phase-coherent with the reference subsystem, up to estimation errors. 
This enables the coherent combination of the subbands across the full band of interest.  

\textbf{Remark.} The proposed method can handle \ac{cfr} estimates obtained at different time instants by the different subsystems, as long as the time difference among the estimates is short enough to consider the channel parameters to be constant. To see this, consider \eq{eq:subband-cfr-carrier} and two \ac{cfr} estimates obtained by subsystems $1$ and $2$ at times $t_1$ and $t_2$, respectively. If the two estimates are sufficiently close in time, the channel parameters $\alpha_l$ and $\tau_l$ can be considered constant. Conversely, the offsets $\tau_{o,2}(t), \phi_{o, 2}(t) $ are fast time-varying, so they change from $t_1$ to $t_2$. Taking subsystem 1 \textit{at time $t_1$}, our approach compensates for the cumulative \ac{to} and \ac{po}, given by  $\tau_{o,2}(t_2) + \tau_{o,2}(t_1)$ and $ \phi_{o, 2}(t_2) + \phi_{o, 2}(t_1)$. 
\name{} is therefore general enough to handle relative \ac{to} and \ac{po} due to collecting \ac{cfr} at different time instants.
This feature makes it extremely flexible in utilizing the \ac{cfr} estimates obtained by the communication system.
Moreover, compared to methods based on linear fitting of the unwrapped phase at different subcarriers, e.g., \cite{kotaru2015spotfi, xiong2015tonetrack}, our approach is more effective with non-contiguous frequency bands (see \secref{sec:results}).

\subsection{Multiband fusion}\label{sec:omp}

Once the available subbands have been made mutually coherent, we use \ac{omp} to obtain a combined set of delay estimates. This step is similar to the intra-subsystem \ac{cfr} reconstruction in \secref{sec:intra-subsys}. However, the search space is huge since the number of subcarriers in the full band is much larger than that in the single subsystems, i.e., $K \gg K_i$.  
Hence, a direct application of \ac{omp} would either have prohibitive computational complexity if we select a small grid spacing, or give inaccurate results if the grid spacing is too large. To solve this issue, we leverage the knowledge of the delays estimated by the single subsystems before the coherent fusion to greatly reduce the search space. This can be thought of as focusing the \ac{omp} algorithm around the solutions obtained from the lower-resolution subsystems.

\textbf{Search space reduction.} To effectively reduce the search space, we consider the reference subsystem and denote by $\{\tau_{1,1}, \dots, \tau_{1,L_1^{\rm OMP}}\}$ its set of delays. Instead of considering a complete grid of delays, as done in the intra-subsystem \ac{cfr} reconstruction, here \name{} works on a union of discrete intervals around the reference subsystem's solutions.
To do so, we first obtain a continuous union of intervals around the candidate delays from the reference subsystem, i.e., $\mathcal{R} = \bigcup_l \mathcal{R}_l$, where $\mathcal{R}_l = [\tau_{1,l} - \gamma , \tau_{1,l} + \gamma ]$. $\gamma $ is the one-sided width of the interval and it is chosen as half the worst nominal delay resolution among the subsystems, i.e., $\gamma  = 1/(2\min_i(B_i))$. This way, the search area around the subsystem's solution is conservatively adapted to the subsystem with the lowest resolution.

Then, \name{} discretizes $\mathcal{R}$ to construct a grid of $Q$ candidate delays for \ac{omp} with step $\delta$. The latter determines the final \name{} delay resolution and accuracy. The elements of the discretized set of delays are denoted by $\nu_1, \dots, \nu_Q$.

Thanks to the search space reduction step, the dimensionality of the problem is reduced from having $K$ candidates to $Q$ candidates. $Q$ can be significantly smaller than $K$, since $L_1^{\rm OMP}$ is small due to the sparsity of the channel and $\delta$ is under our control. 

\textbf{\ac{omp}-based solution.} Similar to \secref{sec:intra-subsys}, the subsystem combination can be cast as a compressed sensing problem that we solve using \ac{omp}. To this end, we construct a matrix with the candidate complex sinusoidal components in the channel as its columns. This is a \textit{partial} $K\times Q$ Fourier matrix, $\mathbf{F}$, that only contains the candidates corresponding to the delays in set $\{\nu_1, \dots, \nu_Q\}$. 
The components of $\mathbf{F}$ are defined as
$\left( \mathbf{F}\right)_{k, q} = e^{j2\pi k \nu_{q} \Delta_f} / \sqrt{K}$. 
To model the missing subcarrier measurements, we introduce a selector matrix, $\mathbf{A}\in \{0,1\}^{M\times K}$, whose rows are the $K$-dimensional vectors of all zeros but the $k$-th component, and each row has a different $k \in \{k_{1,1}, \dots, K_{1,1}-1, \dots, k_{C,S}, \dots, K_{C,S}-1\}$. Left-multiplying $\mathbf{A}$ by $\mathbf{F}$ selects the rows of $\mathbf{F}$ whose indices are in the set of available \ac{cfr} samples.
Call $\mathbf{\Gamma} = \mathbf{A}\mathbf{F} \in \mathbb{C}^{M\times Q}$, and define the \ac{cfr} vector of dimension $M$
\begin{equation}
    \bar{\mathbf{H}} =[\bar{H}_{1, 1, 0}, \dots, \bar{H}_{1, 1, K_1}, \dots, \bar{H}_{C, S, 0}, \dots, \bar{H}_{C, S, K_S}]^{\mathsf{T}},
\end{equation}
that contains all the available measurements from all the $S$ subbands after \ac{to} and \ac{po} compensation.
We denote by $\mathbf{h}\in \mathbb{C}^{Q}$ the \ac{cir} obtained by fusing the $S$ subbands. Using \ac{omp} we estimate $\mathbf{h}$ by solving the same problem in \eq{eq:subsystem-cir}, using $\bar{\mathbf{H}}$ as the measurement vector and $\mathbf{\Gamma}$ as the model matrix.
As for the single subsystems, \ac{omp} is stopped once the reconstruction error with respect to the measurements reaches a $5\%$ threshold, and the corresponding number of non-zero components of $\hat{\mathbf{h}}$ is $L^{\rm OMP}$.
The set of path delays and amplitudes obtained from the non-zero components of $\hat{\mathbf{h}}$ is $\{\hat{\alpha}_{l}, \hat{\tau}_{l}\}, l=1,\dots,L^{\rm OMP}$. The delays are then mapped to relative distances as $\hat{r}_l = c \hat{\tau}_l - D$, where $D$ is the distance between the \ac{tx} and the \ac{rx}, assumed known. Relative distances can be used to localize a target in both mono-static and bi-static scenarios, as described, e.g., in \cite{kuschel2019tutorial, pegoraro2023rapid, pegoraro2024jump}.

\subsection{Temporal aggregation}\label{sec:aggregation}

In this section, we discuss how \name{} can improve its ranging accuracy and resolution by aggregating multiple channel estimates across time, as detailed in \alg{alg:aggregation}. 

Consider a sequence of $N$ path delays, $\{\hat{\tau}_{1}(t_n), \dots, \hat{\tau}_{L^{\rm OMP}}(t_n)\}$, and amplitudes, $\{\hat{\alpha}_{1}(t_n), \dots, \hat{\alpha}_{L^{\rm OMP}}(t_n)\}$, for $n=1, \dots, N$, obtained by applying \name{} to different \ac{isac} packets or \ac{ofdm} slots at time instants $t_1, \dots, t_N$. These must be obtained in a short processing interval such that the channel parameters can be considered constant, i.e., \mbox{$t_N - t_1$} should be within the coherence time of the channel. 

Recall that the delays outputted by \ac{omp} belong to a discrete grid of candidates $\nu_q$, with $q= 1, \dots, Q$, which is kept fixed during the aggregation period.
The temporal aggregation step is based on the following observation: When applied within the coherence time of the channel, \name{} outputs correlated sets of delays that can be aggregated (coherently or incoherently). 
To do so, \alg{alg:aggregation} iterates over the elements in the delay grid and over time slots (lines 2-3). Then, if $\nu_q$ is among the set of outputs of \ac{omp} in the considered slot $n$, we accumulate it across time using a running average of the path amplitudes, $\chi_q$ (lines 4-6), that combines the current amplitude $\hat{\alpha}_l(t_n)$ with the previous $\chi_q$.  

We propose two alternative versions of the running average, one for static targets and one for human sensing (line 5), respectively: 
\begin{itemize}
\item If the targets are static, the temporal aggregation can be performed by taking into account the phase of the complex amplitudes (coherent aggregation), which gives higher \ac{snr} and resolution. This corresponds to the first case in line 5 of \alg{alg:aggregation}.
\item With dynamic targets such as humans, instead, we only aggregate the magnitude information for each path (incoherent aggregation) since the time variation of the phase due to respiration or slight movement would lead to destructive interference. This corresponds to the second case in line 5.
\end{itemize}
Finally, in line 9, we select as the final improved set of delays the $L^{\rm OMP}$ candidates for which $\chi_q$ is highest, which we denote by $\bar{\tau}_l$, for $l=1, \dots, L^{\rm OMP}$. The channel gains of such delays are the corresponding $\chi_q$, which we call $\bar{\alpha}_l$. The delays are then mapped to relative distances as $\bar{r}_l = c (\bar{\tau}_l - \bar{\tau}_1)$.

\begin{centering}
\begin{algorithm}[t!]
\footnotesize
	\caption{\name{} multipath temporal aggregation.}
	\label{alg:aggregation}
	\begin{algorithmic}[1]
		\REQUIRE Set of delays and amplitudes across time $\hat{\tau}_{l}(t_n), \hat{\alpha}_{l}(t_n)$, for $l=1, \dots, L^{\rm OMP}$ and $n=1, \dots, N$, \ac{omp} grid $\nu_1, \dots, \nu_Q$, order $L^{\rm OMP}$.
		\ENSURE Improved set of delays and amplitudes $\{\bar{\alpha}_l, \bar{\tau}_l\}, l=1,\dots,L^{\rm OMP}$.
  \STATE Initialize $q = 1$, $\chi_q = 0, \forall q = 1, \dots, Q$.
  \FOR{$q = 1, \dots, Q$}
  \FOR{ $n = 1, \dots, N$}
  \IF{$\nu_q \in \{\hat{\tau}_{l}(t_n)\}, l=1, \dots,L^{\rm OMP}$}
  % \STATE $\chi_q \leftarrow \chi_q + 1/N$
   \STATE $ \chi_q \leftarrow  \left\{
\begin{array}{cc}
 \left[\hat{\alpha}_{l}(t_n) + n \chi_q \right] / (n+1) &\mbox{ if target is static} ,\\
  \left[|\hat{\alpha}_{l}(t_n)| + n \chi_q \right] / (n+1) &   \mbox{otherwise}\nonumber.
\end{array}\right.
$
  \ENDIF
  \ENDFOR
  \ENDFOR
  \STATE Return $\{\bar{\alpha}_l, \bar{\tau}_l\}, l=1,\dots,L^{\rm OMP}$ as the $L^{\rm OMP}$ path delays with the highest $\chi_q$.
	\end{algorithmic}
\end{algorithm}
\end{centering}

\section{Implementation}

\begin{figure}[t!]
	\begin{center}   
		\centering
		\subcaptionbox{\label{fig:testbed}}[0.56\columnwidth]{\includegraphics[width=0.56\columnwidth]{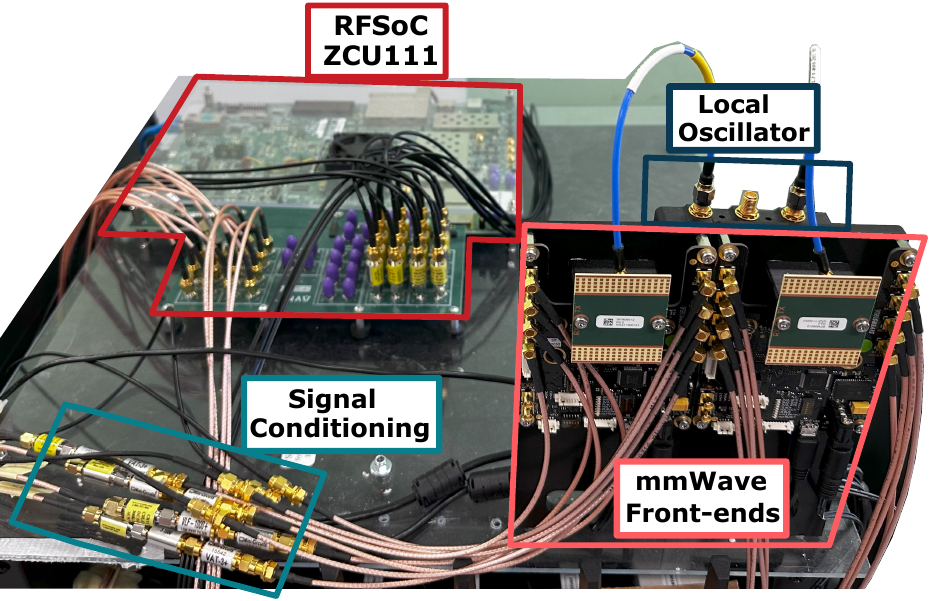}}
		\subcaptionbox{\label{fig:lab}}[0.42\columnwidth]{\includegraphics[width=0.42\columnwidth]{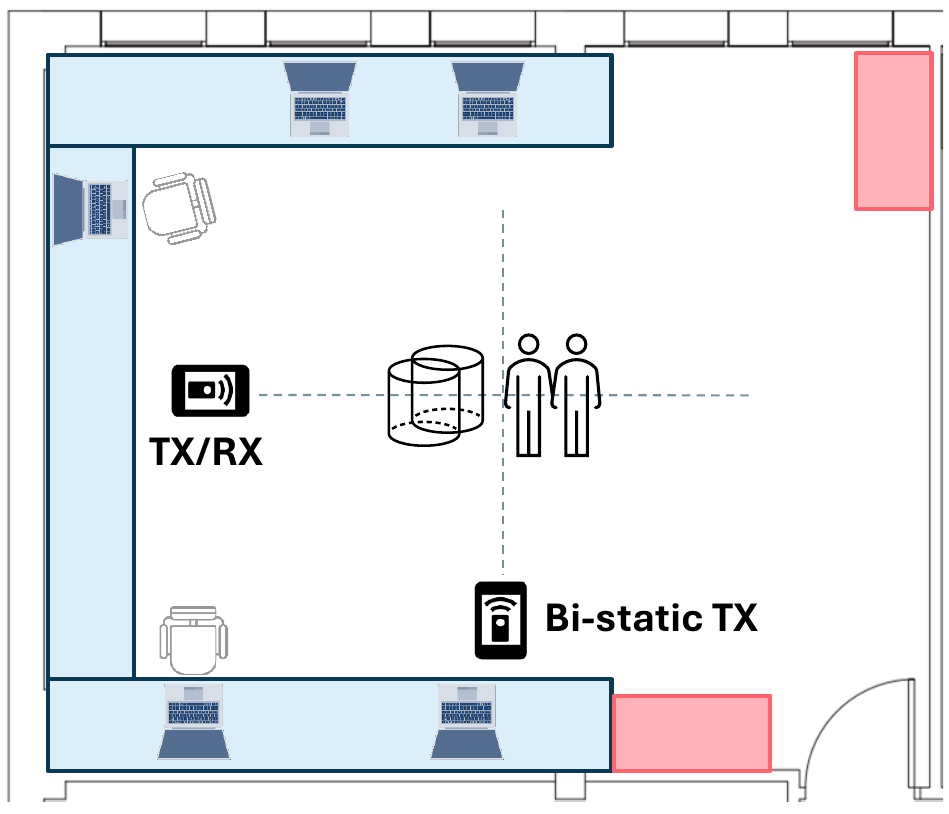}}
		\caption{\name{} prototype (a) and experimental environment (b).}
		\label{fig:setup}
	\end{center}
\end{figure}

To implement \name{}, we use the open-source Mimorph platform \cite{Lacruz_MOBISYS2021} as a baseline. The platform includes an AMD (Xilinx) \ac{rfsoc} that comprises \ac{fpga} logic, multiple analog-to-digital/digital-to-analog converters, and ARM processors. We implement \name{} to work in the unlicensed 58-64~GHz \ac{mmwave} band using linear antenna array front-ends with 16 elements from Sivers Semiconductors, suitable for analog beamforming. We choose a \ac{mmwave} frequency band as it represents a challenging test case, given the high sensitivity to \ac{cfo} and the strong phase noise of high frequencies~\cite{rasekh2019phase}.
Signal conditioning from the \ac{rfsoc} to the \ac{mmwave} front-ends includes DC-block filters, low-pass filters (1~GHz cut-off frequency), and 3~dB attenuators. The main components of a \ac{tx}-\ac{rx} node of the testbed are shown in \fig{fig:testbed}. We configure the testbed to work as $C=2$ incoherent subsystems with carrier frequencies 60.48~GHz and 62.64~GHz.  
The testbed can operate concurrently as \ac{tx} and \ac{rx}, in mono-static configuration, or as a bi-static system, using two nodes like the one in \fig{fig:testbed}.

\textbf{Signal generation.} 
Signals for each sub-system are generated offline. Evaluating \name{} requires collecting full band \ac{cfr} estimates and other information used as ground truth, as detailed in~\secref{sec:exp-base}. 
Therefore, we generate a composite packet including \ac{5g} \ac{ofdm} symbols and IEEE~802.11ay channel estimation fields. 
An \ac{ofdm} symbol including \ac{dm-rs}, spanning the full bandwidth, is used as ground truth (see the \textit{Full band} baseline in \secref{sec:exp-base}).
An IEEE~802.11ay channel estimation field is used for \ac{sc} \ac{cir} estimation.
\ac{5g} pilot signals spanning different subbands are used for \name{}, with bandwidth and starting frequency depending on the specific experiment. 
The signal is sent to the \ac{rfsoc} using an Ethernet port and stored in loopback memories implemented in the \ac{fpga} logic. The \ac{fpga} clock is set to $245.76$~MHz with a super-sampling rate factor of 8, giving an equivalent of $1966.08$~MHz. The inter-packet time is configurable in runtime by a host PC. Since more than one subsystem is employed, independent data paths are used and connected to independent \ac{mmwave} front-ends (\fig{fig:testbed}). 

\textbf{Signal capture and saving.} 
To enable the testbed operation in mono-/bi-static operation, we modify the packet detection block from \cite{Lacruz_MOBISYS2020} to either trigger packet capture when detecting a valid preamble in bi-static operation mode or trigger the capture when transmitting a packet (mono-static mode). The operating mode can be updated at runtime from a host PC. Valid packets are stored in on-board RAM (up to 4GB) and then these are offloaded through a 10~Gb Ethernet interface to be processed offline. 

\section{Experimental results}\label{sec:results}

In this section, we describe the experimental setup used in the evaluation of \name{}. Then we provide an in-depth analysis of our results obtained in different settings. 

\subsection{Experiments and baselines}\label{sec:exp-base}

\textbf{Experiments description.} 
To evaluate \name{}, we perform 27 experiments involving \ac{isac} radio channel measurements in different scenarios and configurations. 
Each experiment is repeated 5 times and involves the transmission of 100 packets with an inter-packet time of 50~ms (unless stated otherwise). 
The experiments are carried out indoors, in a 7~m~ $\times$~6~m room (see \fig{fig:lab}), and can be divided into six groups detailed in the following. In groups (1)-(5) we use metal cylindric reflectors as targets, while group (6) involves human subjects. Groups (1)-(4) and (6) are obtained in a mono-static scenario, while in (5) we use a bi-static setting. In our experiments, we point the TX beam pattern toward the targets, unless stated otherwise. In practice, algorithms such as the one in~\cite{pegoraro2024jump, pegoraro2023rapid} can be used to estimate the angular location of the scatterers.
As an anchor path to achieve phase coherence, we use the self-interference path in the mono-static scenario and the \ac{los} in the bi-static one.

\textit{(1) 2 targets (8 experiments):} 2 metal cylinders are placed at different distances from the system, ranging from 1.5 to 5~m. The inter-target distance is changed from 30~cm to 60~cm.

\textit{(2) 3 targets (5 experiments):} 3 metal cylinders are placed at different distances from the system, ranging from 2 to 5~m. The inter-target distance is changed from 30~cm to 60~cm.

\textit{(3) Resolution limit test (3 experiments):} 2 metal cylinders are placed at 17.2, 10.1, and 3.1~cm inter-target distance, to evaluate the maximum resolution achieved by \name{}. The distance of the second target from the \ac{tx} is 2.78~m.

\textit{(4) Changing angle (5 experiments):} 2 metal cylinders are placed about 2.5~m from the \ac{tx} with 33~cm inter-target distance. We change the angular location of the targets in different experiments among $- 30^{\circ}, -15^{\circ}, 17^{\circ}, 30^{\circ}$. In each, experiment, we change the antenna beam pattern used by the \ac{tx} to point at the targets. This scenario is of high practical interest since pilot signals in \ac{isac} systems (e.g. \acp{ssb}) are often beamformed in different directions.

\textit{(5) Bi-static scenario:}
This group of experiments is performed in a bi-static scenario with a distance of 3.24~m between the \ac{tx} and the \ac{rx}. 2 metal cylinders are placed close to each other so that the segments connecting the \ac{tx} to the target, and the target to the \ac{rx} form a $90^{\circ}$ angle (bi-static angle). The inter-target bi-static distance changes in different experiments from 3.5~cm to 8.9~cm. This scenario is particularly challenging since the nominal ranging resolution in the bi-static case is degraded by a factor of 0.7 due to the $90^{\circ}$ bi-static angle, as shown in \secref{sec:delay-resolution}.

\begin{figure}[t!]
	\begin{center}   
		\centering
        \subcaptionbox{Bi-static setup of subsystems 1 and 2. \label{fig:bistatic-pic}}[0.65\columnwidth]{\includegraphics[width=0.65\columnwidth]{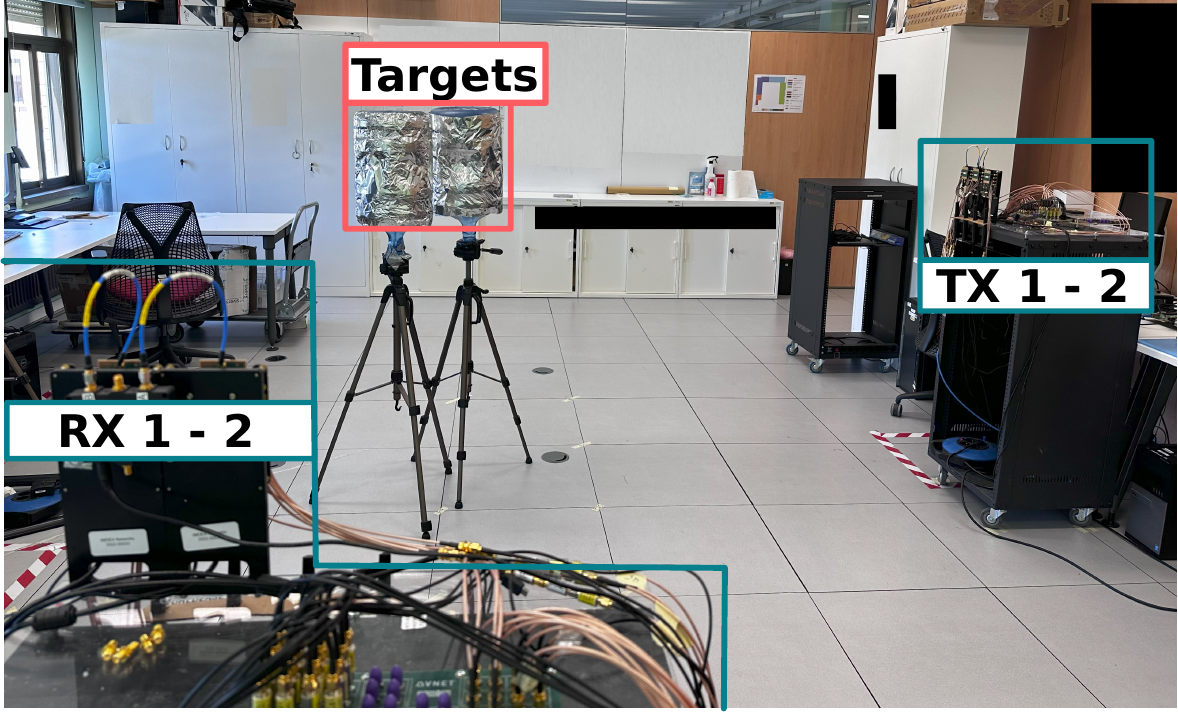}}
		\subcaptionbox{Human sensing. \label{fig:people-pic}}[0.31\columnwidth]{\includegraphics[width=0.31\columnwidth]{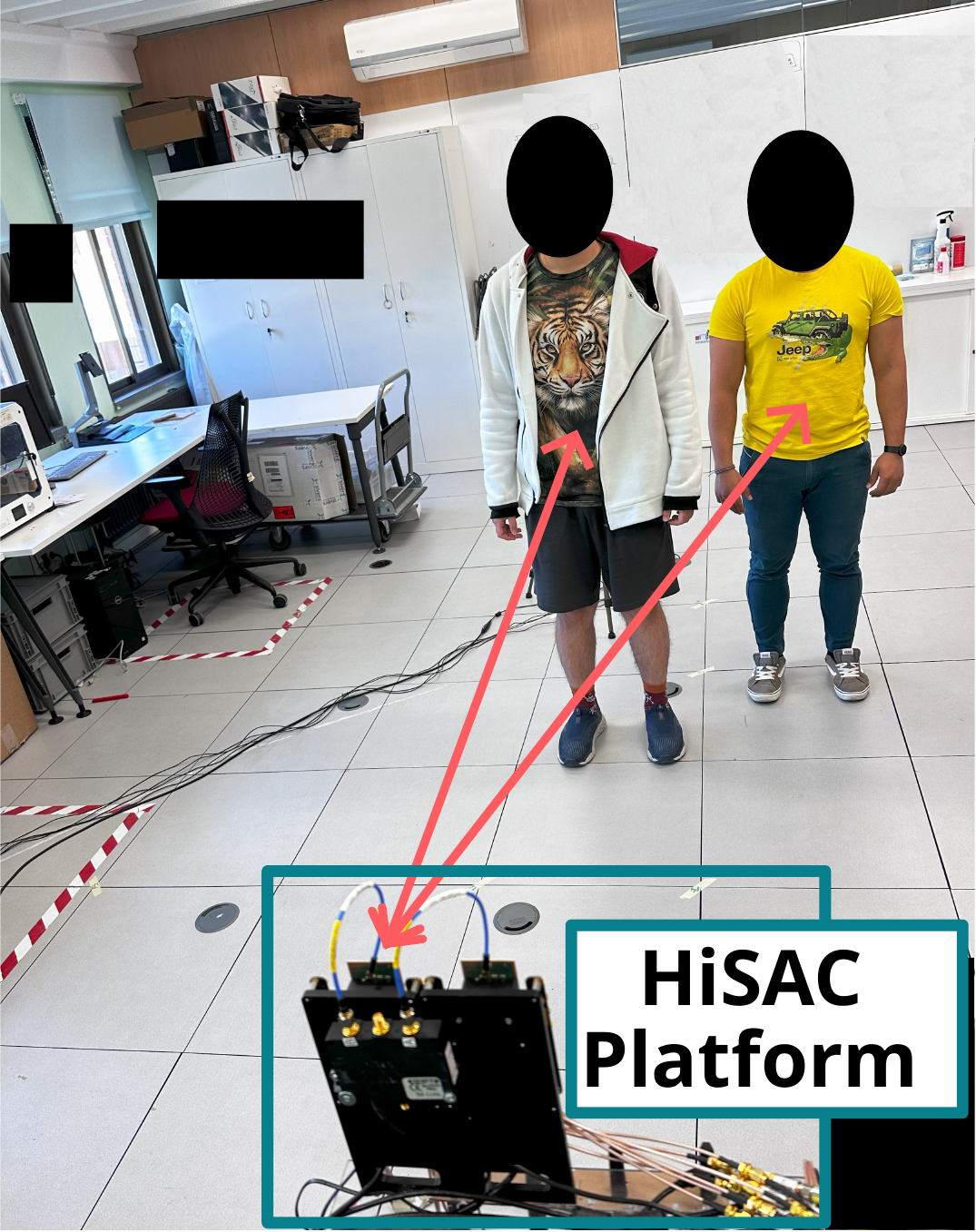}}
		\caption{Experiments in the bi-static setting and with static subjects.}
		\label{fig:bist-people-pics}
	\end{center}
\end{figure}
 
\begin{figure*}[t!]
	\begin{center}   
		\centering
		\subcaptionbox{CA-C1. \label{fig:ca-c1}}[0.9\columnwidth]{\includegraphics[width=0.9\columnwidth]{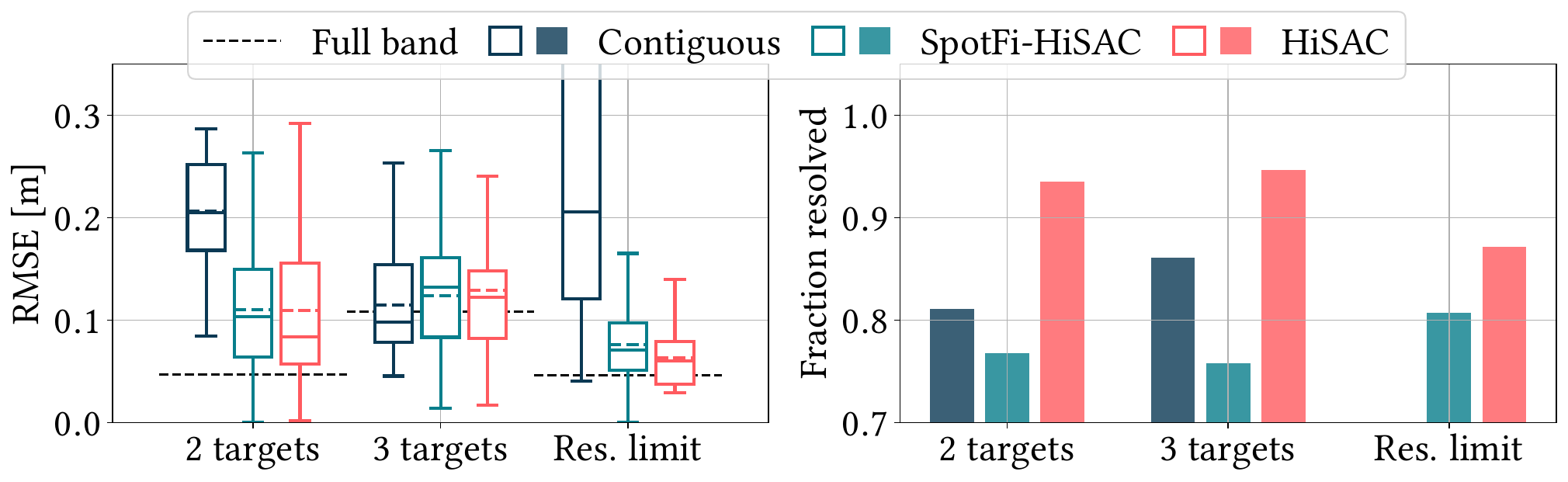}}
		\subcaptionbox{CA-C2. \label{fig:ca-c2}}[0.9\columnwidth]{\includegraphics[width=0.9\columnwidth]{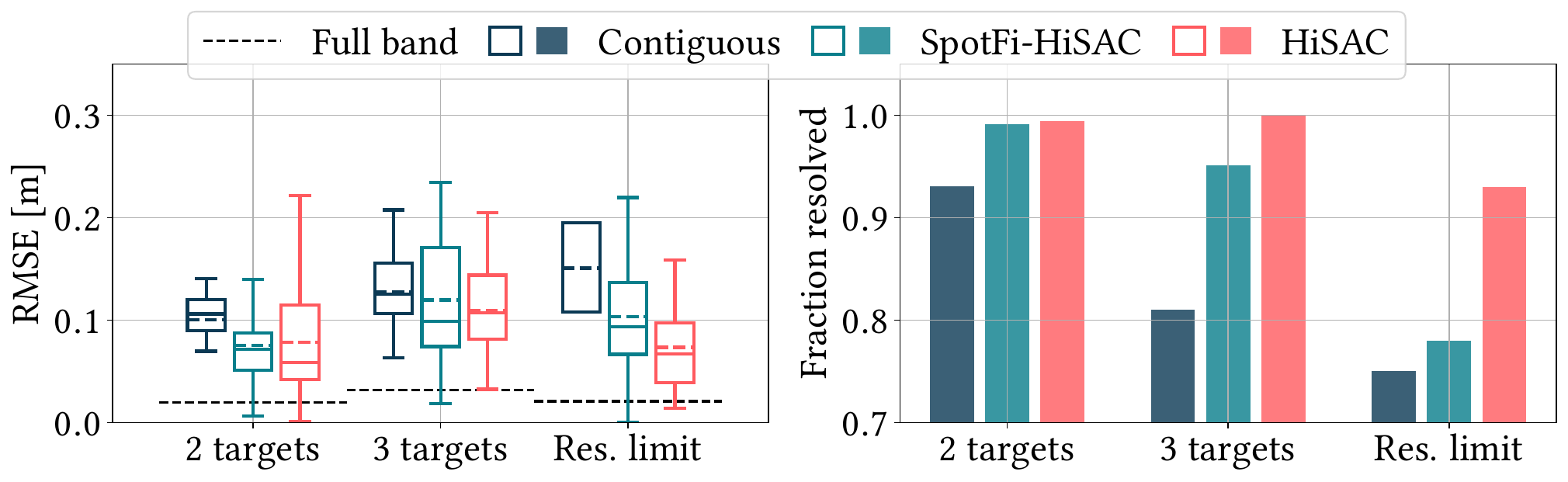}}
		\caption{\name{} results in the CA-C1/C2 setting. We report the ranging \acs{rmse} and the \acs{frt} in experiment groups (1)-(3).}
		\label{fig:ca-tot}
	\end{center}
\end{figure*}
\textit{(6) Human localization and tracking:} A final group of experiments is performed with human targets, to demonstrate the effectiveness of \name{} on weaker multipath components with respect to metal reflectors and its robustness to movement. These involve (i)~2 static subjects standing at 2.30 and 2.64~m from the \ac{tx} as shown in \fig{fig:people-pic}, (ii)~2 moving targets walking back and forth from 3 to 1~m from the \ac{tx} (in this case, the inter-packet time is reduced to 5~ms). Experiments involving people have been carried out in compliance with the IRB of our institute and do not disclose information about the subjects.

\textbf{Baselines for comparison.} 
Since \name{} is the first method to perform multiband \ac{isac}, we design the baseline methods described in the following. Note that~\cite{wang2024uwb}, which is the closest prior work, is not suitable for comparison since (i)~it is based on a deep neural network trained on sub-6~GHz signals so that it would need extensive data collection and retraining on our implementation, (ii)~it uses a channel hopping scheme to collect \ac{cfr} samples that is specific to sub-6 GHz Wi-Fi and does not apply to \ac{5g}.

\textit{Laser telemeter.} We collect ground truth distance measurements of the targets using a laser telemeter, mounted on the \ac{tx}-\ac{rx} antenna front-end.  

\textit{Full band.} We collect \ac{cfr} measurements over an equivalent bandwidth to the full band of interest and use \ac{omp} to obtain target distance estimates. This is a benchmark to assess how close \name{} gets to the performance of a wideband \ac{isac} system with a bandwidth equal to its total virtual bandwidth.  

\textit{Contiguous band.} We collect \ac{cfr} measurements over a \textit{contiguous} region of the spectrum with a bandwidth equal to the real bandwidth of \name{}, which includes only the subcarriers in which the channel is measured. \ac{omp} is then used to obtain distance estimates. 

\textit{SpotFi-based \name{} (SpotFi-\name{}).} To demonstrate the effectiveness of \name{}'s algorithm to achieve phase coherence across subbands, we design a competitor algorithm that uses SpotFi's approach,~\cite{kotaru2015spotfi}, to perform this task. SpotFi first extracts the phase of the \ac{cfr} in each subband, representing it as a function of the subcarriers, and applies phase unwrapping to remove the wrapping around $2\pi$. Due to the linearity of the phase as a function of path delays, as can be seen from \eq{eq:fullband-cfr}, a linear model can be fit to the unwrapped \ac{cfr} of each subband. Then, the phase offsets are removed by applying a linear transformation to the phase of the \acp{cfr}. The parameters of the transformation are obtained by matching the slopes and intercepts of the linear models across subbands.

To fairly compare the effectiveness of the phase offsets compensation algorithm, we keep the rest of the delay estimation process the same as in \name{}. This step is necessary since the original SpotFi uses the \ac{music} algorithm to estimate delays, it can not handle non-contiguous subbands. Hence, SpotFi-HiSAC is an \textit{improvement} of the original SpotFi that can be applied to non-contiguous subbands. Specifically, SpotFi-HiSAC applies, in order: the intra-subsystem \ac{omp}-based reconstruction from \secref{sec:intra-subsys}, the SpotFi phase offsets removal step, based on line fitting, and the multiband fusion step from \secref{sec:omp}. 

\textbf{Evaluation metrics.}
We adopt two main metrics to evaluate \name{}. The first one is the \ac{rmse} in the distance estimation, computed with respect to the laser telemeter distance measurement. 
\ac{rmse} can only be computed for the targets that are detected by the algorithm, and it is undefined for unresolved targets. Therefore, we introduce a second metric which we call \ac{frt}. This represents the fraction of targets that an algorithm can resolve, i.e., detect correctly, with respect to the total number of targets resolved by the full band baseline. 
We consider a target to be correctly detected by an algorithm if this outputs a target distance sufficiently close, i.e., closer than the minimum inter-target distance in the experiment, to the laser telemeter ground truth distance for that target.
The two metrics should be jointly considered in each evaluation since an algorithm may yield a very low \ac{rmse} but have low resolution, which means it is not exploiting the increased bandwidth. Conversely, an algorithm could have high resolution but poor accuracy, giving a high \ac{rmse}.

\subsection{In-depth evaluation}\label{sec:indepth-eval}
In this section, we evaluate \name{} in the three use cases from \secref{sec:use-cases}, using the experiments from \secref{sec:exp-base}. 

\begin{figure*}[t!]
	\begin{center}   
		\centering
		\subcaptionbox{BWP-C1. \label{fig:bwp-c1}}[0.9\columnwidth]{\includegraphics[width=0.9\columnwidth]{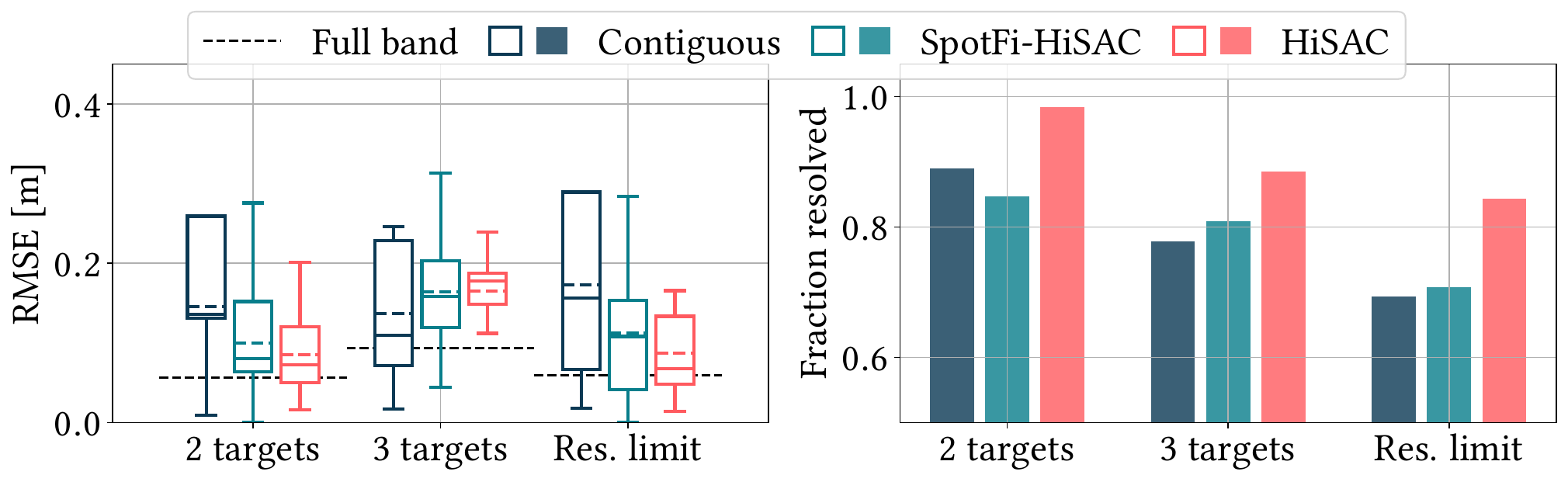}}
		\subcaptionbox{BWP-C2. \label{fig:bwp-c2}}[0.9\columnwidth]{\includegraphics[width=0.9\columnwidth]{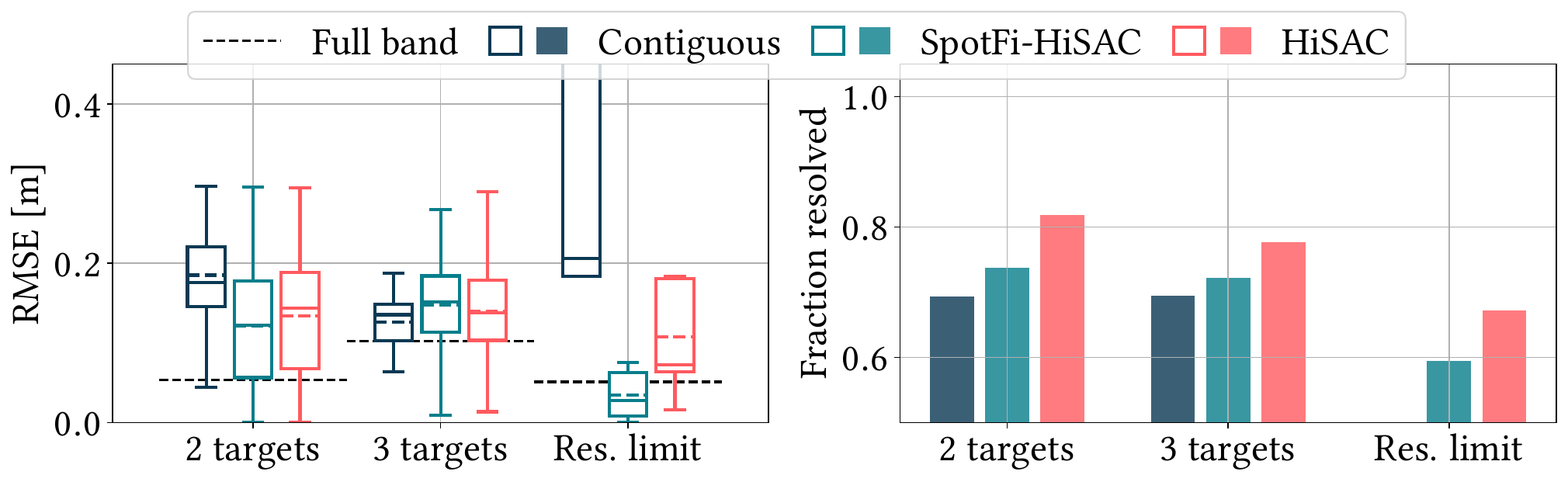}}
		\caption{\name{} results in the BWP-C1/C2 setting. We report the ranging \acs{rmse} and the \acs{frt} in experiment groups (1)-(3).}
		\label{fig:bwp-tot}
	\end{center}
\end{figure*}

\textbf{Carrier Aggregation.} We start our evaluation with the carrier aggregation scenario, combining multiple \ac{5g} channels with $100$ and $400$~MHz bandwidth. We assume that in this case the \ac{cfr} is estimated over the full channels, similar to using \ac{dm-rs} pilots that span the full bandwidth~\cite{5G-NR_R15}.
We consider two configurations with 2 subsystems: 

(i)~Configuration 1 (CA-C1), including 4 subbands with an effective bandwidth of 400~MHz and a virtual bandwidth of 2.01~GHz. The first two subbands belong to subsystem 1, while the second two to subsystem 2. The starting frequencies of the subbands relative to the first one are $\{0, 0.19, 1.63, 1.91\}$~GHz, while their bandwidths are all equal to 100~MHz;

(ii)~Configuration 2 (CA-C2), including 5 subbands with an effective bandwidth of 800~MHz and a virtual bandwidth of 3.46~GHz. The first three subbands belong to subsystem 1, while the second two to subsystem 2. The starting frequencies of the subbands relative to the first one are $\{0, 0.19, 1.2, 2.88, 3.36\}$~GHz, while their bandwidths are $\{0.1, 0.1, 0.4, 0.1, 0.1\}$~GHz.

The main challenge in the carrier aggregation scenario is to effectively combine subbands that are widely separated in the spectrum. \fig{fig:ca-tot} shows the \ac{rmse} and \ac{frt} obtained by \name{} in experiment groups (1)-(3) with the carrier aggregation use case. The horizontal dashed line represents the average \ac{rmse} obtained using the full band \ac{cfr}. With CA-C1, \name{} achieves accurate ranging with an average \ac{rmse} below 15~cm in all the experiments. The case with 3 targets gives the highest average \ac{rmse} due to the higher complexity of the multipath environment. 
The contiguous bandwidth and SpotFi-\name{} yield comparable or worse \ac{rmse}. This proves that \name{} gains ranging accuracy thanks to the increased virtual bandwidth and outperforms SpotFi's method to achieve phase coherence.
In terms of \ac{frt}, \name{} provides significant gains. In the resolution limit test, the contiguous band is unable to resolve the targets (0.5 \ac{frt}), while \name{} gives 0.93 \ac{frt}. 
With CA-C2, the overall performance in terms of \ac{rmse} and \ac{frt} on 2 and 3 targets improves for all methods due to the wider virtual bandwidth. However, only \name{} significantly benefits from such increased virtual bandwidth when considering the resolution limit test, achieving higher \ac{frt} compared to CA-C1, while other methods perform slightly worse.

\textbf{Bandwidth Part.} Next, we analyze a bandwidth part scenario, where we combine multiple \ac{cfr} estimates obtained by a \ac{5g} system using \ac{ssb} signals, spanning 240 subcarriers in the middle of the operating channel. Note that here the bandwidth of each subband is significantly lower than in the carrier aggregation case since the \ac{cfr} is measured only on a fraction of the total communication channel. Specifically, using 240~KHz subcarrier spacing, which is reasonable at \ac{mmwave} frequencies, the bandwidth of each \ac{ssb} is $B_{i,s} = 57.6$~MHz.
We consider two different configurations with 2 subsystems: 

(i)~Configuration 1 (BWP-C1), including 4 subbands with an effective bandwidth of 460.8~MHz and a virtual bandwidth of 1.267~GHz. The first four subbands belong to subsystem 1, while the second four to subsystem 2. The starting frequencies of the subbands relative to $f_0$ are $\{0.02, 0.12, 0.22, 0.32, 0.91, 1.01, 1.11,1.21\}$~GHz;

(ii)~Configuration 2 (BWP-C2), including 5 subbands with an effective bandwidth of 230.4~MHz and a virtual bandwidth of 1.267~GHz. The first two subbands belong to subsystem 1, while the second two to subsystem 2. The starting frequencies of the subbands relative to the start of the full band are $\{0.02, 0.12, 0.91, 1.01\}$~GHz.

\begin{figure}[t!]
    \centering
    \includegraphics[width=0.85\linewidth]{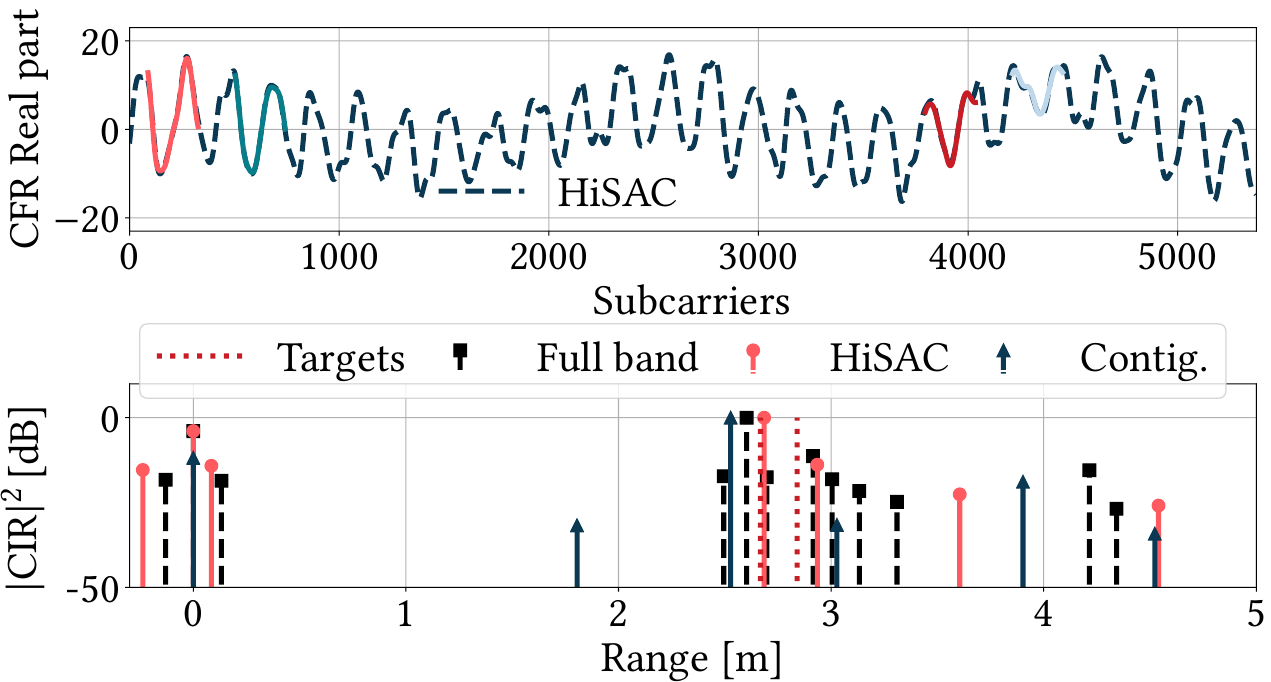}
    \caption{Example \name{} results in BWP-C2. The top plot shows the subbands after coherency has been achieved and the reconstructed \name{} \ac{cfr}. The bottom plot shows the \ac{cir} squared magnitude.}
    \label{fig:example}
\end{figure}

We stress that, as required by bandwidth part operation, each channel estimate in different subbands is obtained at a different time instant, with an inter-packet time of 50~ms. Hence, the spectrum occupied at each time instant is just $57.6$~MHz. An example result on BWP-C1 is shown in \fig{fig:example}, reporting the \ac{cfr} (above) and the squared magnitude of the \ac{cir} (below).
The main challenge in the bandwidth part scenario is to combine narrow available subbands due to the use of unmodified \ac{ssb} pilot signals. \fig{fig:bwp-tot} shows the \ac{rmse} and \ac{frt} obtained by \name{} in the experiment groups (1)-(3) with the bandwidth part use case. With BWP-C1, \name{} outperforms the other approaches. Note that with 3 targets it achieves a slightly higher mean ranging error with respect to the contiguous band (2 cm), but resolves a higher fraction of targets by 0.1.
BWP-C2 represents a very challenging scenario due to the sparsity of the available spectrum (amounting to 18\% of the virtual bandwidth) and to the narrow subbands used. In the resolution limit test, the contiguous \ac{cfr} method fails in resolving the targets and has a large ranging error, while \name{} achieves 11~cm average error and resolves 67\% of the targets on average, with a resolution gain of 13\% over SpotFi-\name{}.

\begin{figure}[t!]
    \centering
    \includegraphics[width=0.9\linewidth]{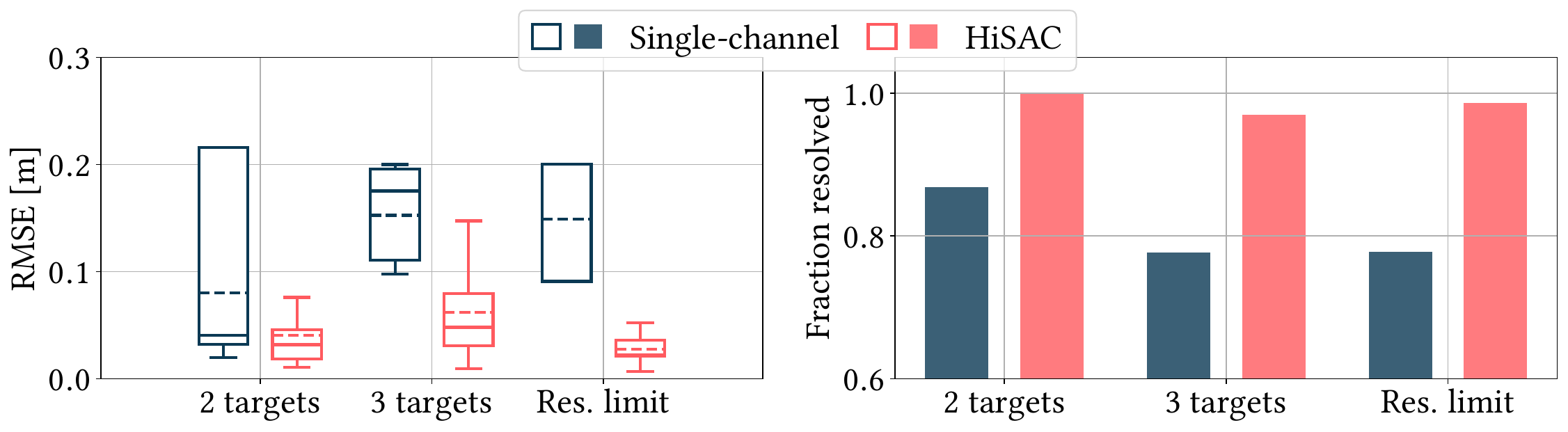}
    \caption{Cross-technology \name{} results.}
    \label{fig:5g-ay}
\end{figure}

\textbf{Cross-technology evaluation.} 
To demonstrate the flexibility of \name{}, we evaluate it in a cross-technology scenario where we combine a 400~MHz \ac{5g} \ac{ofdm} channel, with carrier frequency 59.69 GHz, with a 1.76 GHz IEEE~802.11ay \ac{sc} channel (WiGig) with carrier frequency 62.64 GHz. The virtual bandwidth in this configuration is 4.03~GHz, while the available one is 2.16~GHz.
Note that, in \ac{sc} systems, the \ac{cir} is estimated directly via cross-correlation with the transmitted pilot signals.  
Therefore, before applying \name{}, we convert the IEEE~802.11ay \ac{cir} into the \ac{cfr} using a \ac{dft}.
In \fig{fig:5g-ay}, we show the \ac{rmse} and \ac{frt} results in the cross-technology scenario. As a comparison, we use the targets detected by the peaks of the \ac{cir} estimated by a single IEEE 802.11ay channel. \name{} obtains extremely accurate ranging with an average \ac{rmse} of at most 5~cm (3 targets), whereas the single channel has a worst-case \ac{rmse} of 15~cm. In terms of \ac{frt}, \name{} gives an almost identical resolution to the full bandwidth, with a worst-case \ac{frt} of 0.95 with three targets.

\begin{figure}
    \centering
    \includegraphics[width=\linewidth]{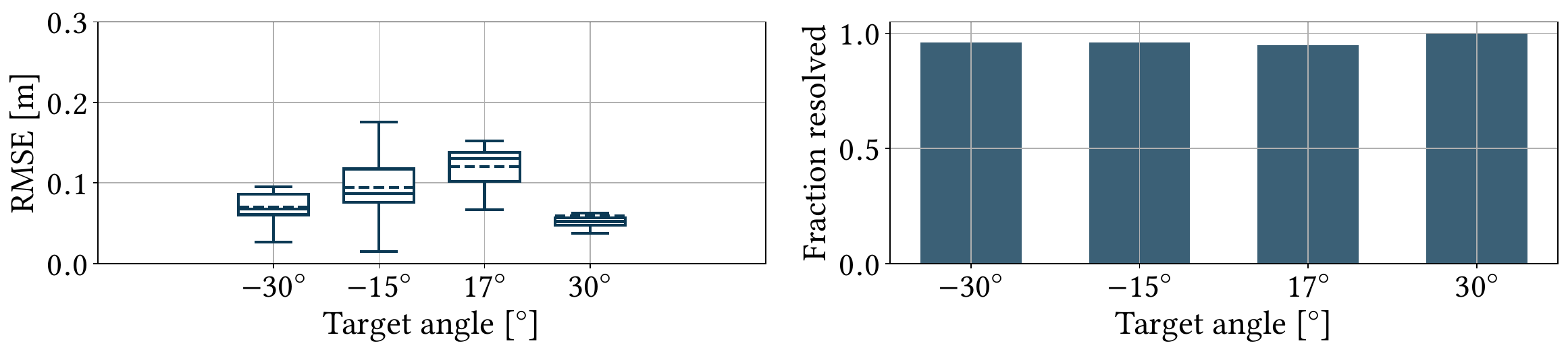}
    \caption{\name{} results changing target angle and transmission beam pattern.}
    \label{fig:angles}
\end{figure}

\textbf{Impact of angle and beamforming.} \ac{isac} systems apply beamforming to direct the signal towards the communication \ac{rx} or targets. \name{} is robust to such changes in the direction of the transmission, as demonstrated by our results in \fig{fig:angles}, obtained on experiments group (4). The ranging \ac{rmse} remains below 15~cm when changing the angle in $[-30^{\circ}, 30^{\circ}]$. The \ac{frt} does not change significantly when changing the transmission angle, proving that \name{} accurately measures distances in different directions.

\begin{figure}[t!]
    \centering
    \includegraphics[width=0.9\linewidth]{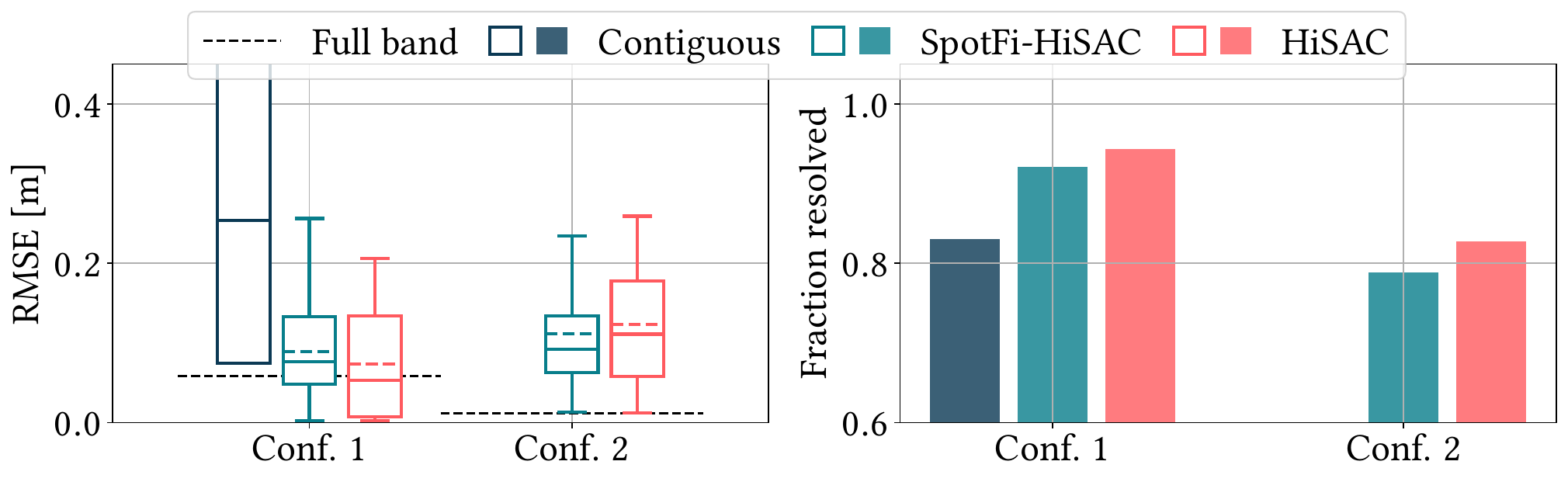}
    \caption{\name{} results in a bi-static bandwidth part scenario.}
    \label{fig:bist}
\end{figure}

\textbf{Bi-static setting.} We evaluate \name{}'s capability to estimate the targets' distances in a bi-static setting, using experiments from group (5) and the BWP-C1/C2 configurations. Note that the full band range resolution in this case is reduced to $\Delta r = c/(2 B \cos(\pi/4))\\\approx 17.5$~cm due to the bi-static angle being 90$^{\circ}$.
\fig{fig:bist} shows the \ac{rmse} and \ac{frt} obtained by \name{} in the bi-static setting. These results are comparable to those obtained in the mono-static setting with a similar configuration, proving that \name{} works when the \ac{tx} and \ac{rx} are widely separated. Conversely, the contiguous \ac{cfr} can not resolve the targets with BWP-C2.

\begin{figure}[t!]
    \centering
    \includegraphics[width=0.9\linewidth]{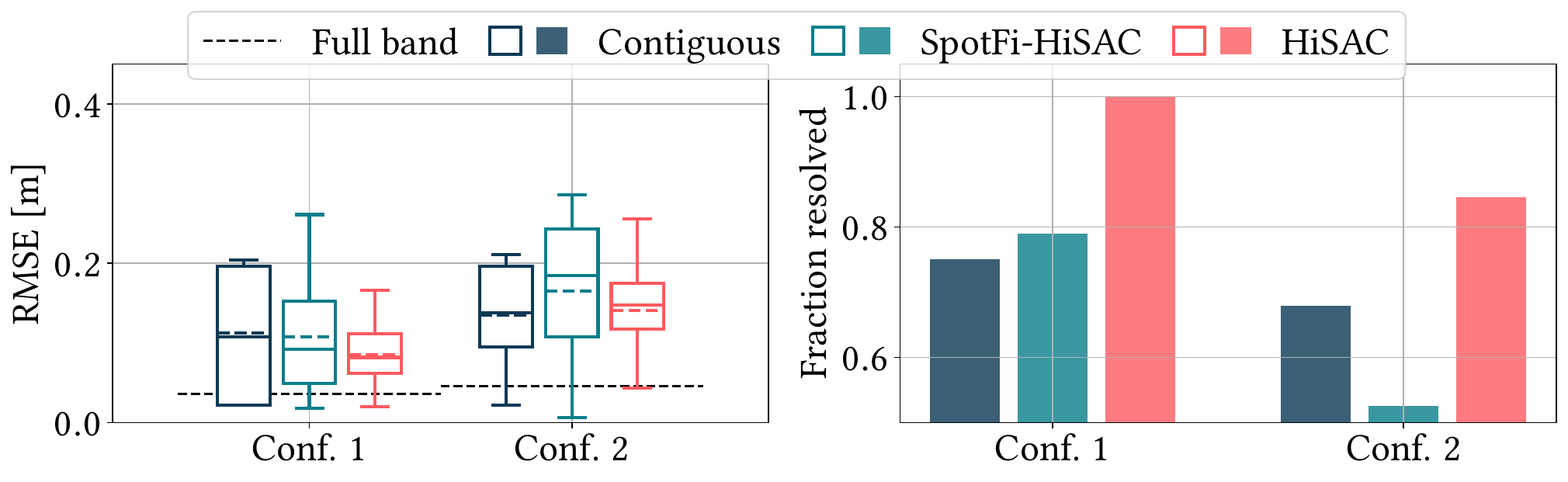}
    \caption{\name{} results with human subjects using bandwidth part.}
    \label{fig:human_static}
\end{figure}

\textbf{People localization.} We test \name{} on human subjects to demonstrate its capability to resolve weaker reflections, using the experiments from group (6).
\fig{fig:human_static} shows the \ac{rmse} and the \ac{frt}. 
These results are obtained using the BWP-C1 and C2 configurations. \name{} achieves 8 and 15~cm average \ac{rmse} with C1 and C2, respectively. Moreover, the \ac{frt} gain that it provides with respect to using the contiguous \ac{cfr} is large: 0.25 using C1 and 0.17 using C2. Note that SpotFi-\name{}'s \ac{frt} with human subjects degrades much more than \name{}'s. Being based on an anchor path, our method to achieve phase coherence is \textit{independent} of the strength of the target multipath component. 

\begin{figure}[t!]
	\begin{center}   
		\centering
		\includegraphics[width=0.45\columnwidth]{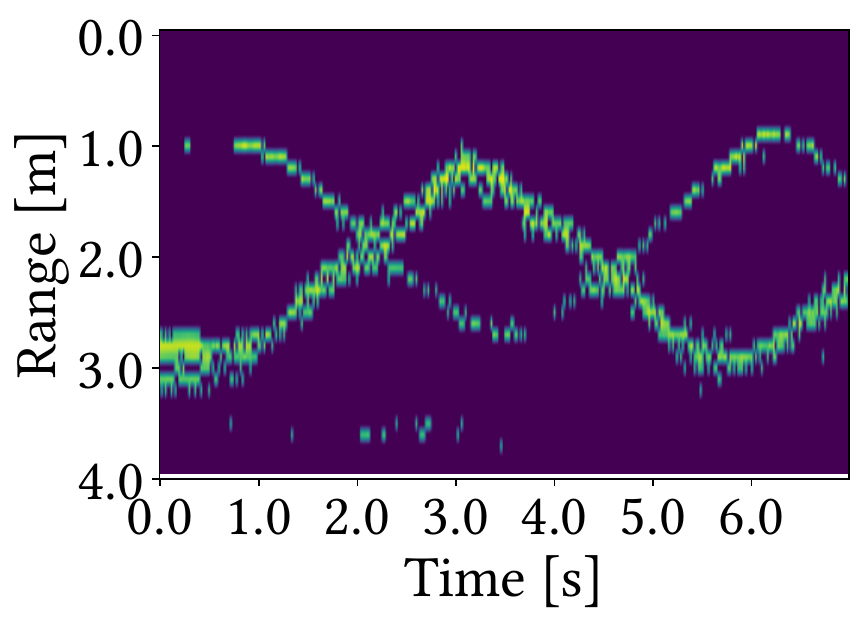}
		\includegraphics[width=0.45\columnwidth]{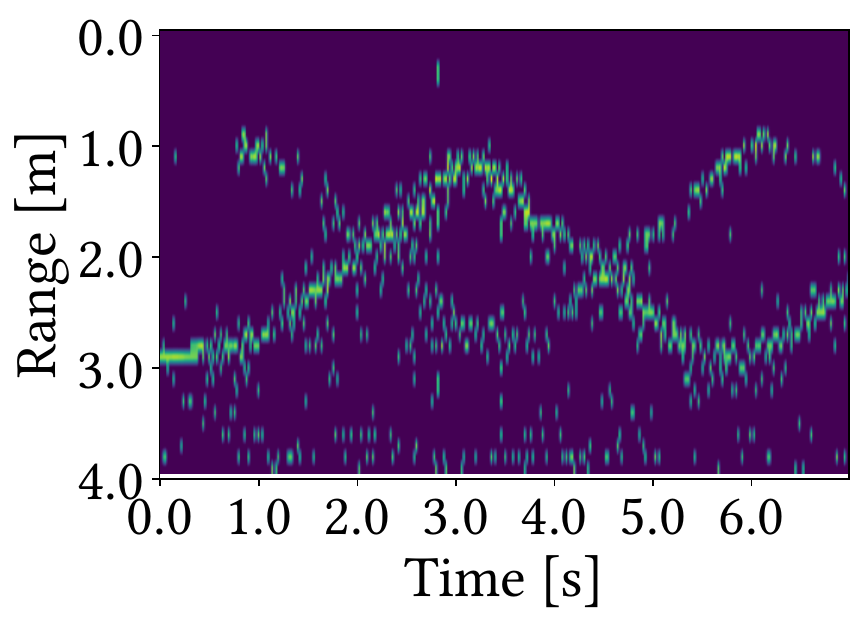}
		\caption{Tracking 2 subjects across time in the BWP-C1 scenario. We plot the normalized \ac{cir} squared magnitude obtained from the reconstructed \ac{cir} with the full band (left) and \name{} (right). \name{}'s resolution is comparable to that of the full 1.2 GHz bandwidth.}
		\label{fig:people-tracking}
	\end{center}
\end{figure}

\textbf{People tracking.} \fig{fig:people-tracking} shows the \ac{cir} squared magnitude across time obtained from the \ac{cir} estimated by \name{} and with the full band with 2 subjects walking back and forth. Peaks in the \ac{cir} (in yellow) correspond to the subjects. Although \name{}'s \ac{cir} is noisier than that of the full band, its resolution, i.e., the capability of distinguishing the two subjects across time, is comparable. The bandwidth used by \name{} in each time slot is only 57.6 MHz (20 times lower than the full bandwidth), i.e., 2.6~m resolution. 

\begin{figure}[t!]
	\begin{center}   
		\centering
\includegraphics[width=0.45\columnwidth]{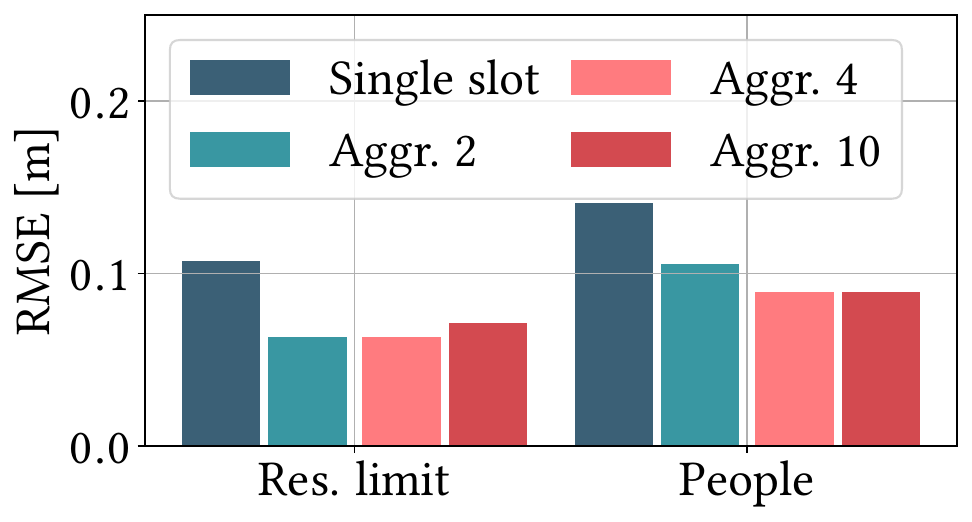}
\includegraphics[width=0.45\columnwidth]{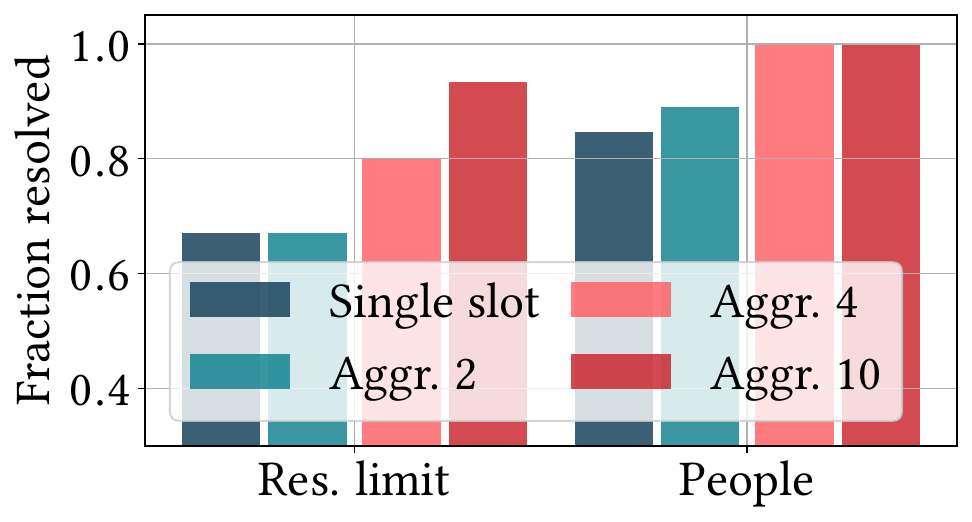}
		\caption{Improvement due to the temporal aggregation (\alg{alg:aggregation}) on experiments (3), Resolution limit, and (6), People, using BWP-C2.}
		\label{fig:time-aggr-tot}
	\end{center}
\end{figure}

\textbf{Temporal aggregation improvement.} \fig{fig:time-aggr-tot} shows the \ac{rmse} and \ac{frt} results obtained by aggregating \name{}'s range estimates over 2, 4, or 10 time slots using \alg{alg:aggregation}. This evaluation is carried out on the most challenging scenarios we considered during the evaluation, namely BWP-C2 and groups of experiments (3) and (6). Our results demonstrate the effectiveness of temporal aggregation, which improves \ac{rmse} by around $30\%$ and \ac{frt} by up to $0.25$ with respect to single-slot \name{}.

\section{Related Work}\label{sec:rel-work}

\textbf{Super-resolution wireless sensing.} Subspace-based super-resolution methods~\cite{kung1983state} and compressed sensing~\cite{eldar2012compressed} are widely used in radar and \ac{isac}, but are still limited by the transmission bandwidth. Some of their recent applications to \ac{isac} are found in~\cite{liu2020super, zhang2023integrated}. Other approaches have employed the spatial diversity of the \ac{rx} array~\cite{zhang2021mmeye} to perform high-resolution imaging in the \ac{mmwave} band, but they can not be applied to low-cost systems with a single \ac{rf} chain such as analog beamforming systems. 
In~\cite{tagliaferri2024cooperative, manzoni2023wavefield}, a novel approach is introduced to apply distributed \ac{sar} techniques to vehicular sensing systems. This approach boosts the resolution by combining distributed devices, thus requiring multiple cooperating nodes. \name{} instead \textit{reuses} the frequency diversity of communication systems without additional requirements. Notably, \name{} could be combined with any of the above techniques to enhance their resolution.

\textbf{Multiband radar sensing.} Multiband sensing has been studied in the radar literature. \cite{cuomo1999ultrawide} has first proposed bandwidth interpolation between two subbands to increase ranging resolution via \ac{ar} modeling and non-linear optimization. Other works have followed a similar research direction adopting different algorithms for combining the subbands~\cite{zhang2014coherent,hussain2018auto, xiong2017coherent,jia2006new,tian2013multiband, zhang2017multiple,van2010high}. 
The above works are based on radar systems with optimized chirp waveforms using wide individual subbands. This significantly simplifies the problem with respect to an \ac{isac} setting where \ac{cfr} estimates are not under control and can be very narrow, making radar approaches underperform. To solve this problem, \name{} innovates with a progressive combination of subbands over coherent subsystems first, and then over the full band of interest.

\textbf{OFDM-based multiband processing.} Several works have demonstrated that combining multiple frequency bands can boost the resolution of \ac{ofdm} systems in active localization.
In \cite{kotaru2015spotfi, xie2015precise, xiong2015tonetrack}, SpotFi, Splicer, and ToneTrack are presented, which combine (\textit{stitch}) multiple contiguous or overlapping Wi-Fi subbands to increase the multipath resolution. They use linear fitting of the unwrapped phase to eliminate phase offsets, which leads to errors when used on non-contiguous, narrow subbands spanning several GHz. 
Other Wi-Fi-based approaches \cite{khalilsarai2019wifi, khalilsarai2020wifi, vasisht2016decimeter} have tackled the same problem but rely on a handshaking process between \ac{tx} and \ac{rx} to eliminate phase offsets, which does not apply to passive sensing in \ac{isac}. 
Similarly, also the frequency hopping strategies used in \cite{bansal2021owll, chen2019m3} require the cooperation of the receiver node, which may not be available in \ac{isac}.
\cite{noschese2020multi, wan2022fundamental, wan2023multiband, kazaz2021delay} propose alternative algorithms based on maximum-likelihood. All these approaches target active localization, in which the \ac{rx} device is the localized target. \name{} instead localizes targets from backscattered reflections, as done by radars.

More recently, \cite{wan2024ofdm} has brought the research attention to exploiting multiband \ac{cfr} to perform wideband radar-like ranging in \ac{isac}. To the best of our knowledge, the only system that tackles this problem is UWB-Fi \cite{wang2024uwb}, in the sub-6 GHz unlicensed spectrum. However, this system is based on a neural network that learns to combine subbands and compensate for phase offsets. Changing hardware, frequency band, or technology (\ac{ofdm} vs \ac{sc}) may require retraining the system which is time-consuming and requires new data. Conversely, \name{} does not require training and generalizes to different implementations.

\textbf{Cross-band channel prediction.} A recent line of work has investigated channel prediction in one frequency band from an available channel estimate in a different band \cite{bakshi2019fast,cho2023scalable,li2023ca,liu2021fire,banerjee2024horcrux,shepard2012argos}. This is typically done to avoid having to transmit feedback channel information in downlink/uplink, thus reducing overhead. Although the general idea of this problem is linked to the multiband setting, predicting the channel in a different band is significantly different from combining multiple non-contiguous incoherent \ac{cfr} estimates to increase the total system bandwidth. Moreover, the above works focus on communication, hence the channel prediction is mostly aimed at estimating \ac{snr} at the \ac{rx}. Conversely, \name{} is an \ac{isac} system that estimates fine-grained complex amplitudes and delays of individual paths in the channel.

\section{Discussion and limitations}\label{sec:discussion}

\textbf{Sensing over very large bandwidth.} \name{} can aggregate subbands over regions of the spectrum spanning several GHz. However, aggregation between very far frequency bands (e.g., sub-6~GHz and \ac{mmwave}) is not feasible, since the frequency-dependency of the scattering coefficients of the different paths would become non-negligible and prevent coherent aggregation. Future research in this direction is key to truly exploit the multiband potential.

\textbf{Impact of the scattering angle.} The phase of the scattering coefficients $\alpha_l(t)$ in \eq{eq:subband-cfr-carrier} is assumed to be approximately constant for all subsystems. This assumption holds for isotropic targets or if the scattering angle is similar across subsystems~\cite{manzoni2023wavefield}. This assumption is reasonable for \ac{isac} \acp{bs} and \acp{ap} that are typically clustered on the same antenna poles. Combining spatially diverse subsystems across multiple bands will be investigated in future work.

\textbf{Impact of narrow subbands.} In the limit case in which each \name{} subsystem estimates a single very narrow subband, reconstructing a subsystem-wide \ac{cfr} model could be challenging. If targets are too close to the \ac{tx}, the \ac{cfr} may not oscillate fast enough to have sufficient information about the path delay in the narrow subband. This is a challenging problem that requires further investigation. 

\textbf{Impact of Doppler.} The impact of Doppler on subbands combination remains unexplored in \ac{isac}. Due to its variation with carrier frequency, Doppler introduces a path- and subsystem-specific phase shift on the \ac{cfr} which may degrade the subbands combination. \name{} is sufficiently robust to this phenomenon as shown in \secref{sec:indepth-eval}. However, further investigation of this aspect is required.

\section{Conclusion}
The problem of achieving range super-resolution in \ac{isac} systems with narrow and discontinuous subbands is addressed in this paper. To solve it, we present \name{}, a general signal-processing method for coherent multiband ranging that enhances the resolution of existing communication systems by only reusing channel estimates obtained via pilot signals. Our approach does not rely on specific hardware or protocol and works across different communication technologies (i.e., \ac{5g} or IEEE 802.11ay). 
Our extensive experiments with objects and humans demonstrate that \name{} enhances the resolution by up to 20 times compared to single-band processing, occupying the same bandwidth in each time slot.

\begin{acks}
The work of Jacopo Pegoraro and Michele Rossi was funded by the European Union under the Italian National Recovery and Resilience Plan (NRRP) Mission 4, Component 2, Investment 1.3, CUP C93C22005250001, partnership on “Telecommunications of the Future” (PE00000001 - program “RESTART”).
The work of Jesus O. Lacruz and Joerg Widmer has been partially funded by project PID2022-136769NB-I00 funded by MCIN/AEI /10.13039/501100011033 / FEDER, UE, and by project RISC-6G TSI-063000-2021-59, funded by Spanish Ministry of Economic Affairs and Digital Transformation, European Union NextGeneration-EU.
\end{acks}

\bibliographystyle{ACM-Reference-Format}
\bibliography{references.bib}

\end{document}